\documentclass[fleqn,usenatbib]{mnras}

\usepackage{newtxtext,newtxmath}

\usepackage[T1]{fontenc}
\usepackage{ae,aecompl}

\usepackage{graphicx}	
\usepackage{amsmath}	
\usepackage{amssymb}	

\usepackage{float}
\usepackage{aliascnt}
\newaliascnt{eqfloat}{equation}
\newfloat{eqfloat}{h}{eqflts}
\floatname{eqfloat}{Equation}

\newcommand*{\ORGeqfloat}{}
\let\ORGeqfloat\eqfloat
\def\eqfloat{%
  \let\ORIGINALcaption\caption
  \def\caption{%
    \addtocounter{equation}{-1}%
    \ORIGINALcaption
  }%
  \ORGeqfloat
}

\title[OVI-PI vs. OVI-CI: redshift, mass and radial dependence]{CGM properties in VELA and NIHAO simulations; the OVI ionization mechanism: dependence on redshift, halo mass and radius}

\author[S. Roca-Fabrega et al.]{
S. Roca-F\`abrega$^{1,2}$\thanks{E-mail: sroca01@ucm.es}, A. Dekel$^{2}$, Y. Faerman$^{2}$, O. Gnat$^{2}$, C. Strawn$^{3}$, D. Ceverino$^{4}$,\newauthor J. Primack$^{3}$, A.~V. Macci\`o$^{5,6}$, A.~A. Dutton$^{5}$, J.~X. Prochaska$^{7,8}$, J. Stern$^{9}$
\\
$^1$ Dept. F\'isica de la Tierra y Astrof\'isica, Facultad de Ciencias F\'isicas, Instituto de F\'isica de Part\'iculas y del Cosmos,\\ Universidad Complutense de Madrid, Plaza de las Ciencias 1, Madrid, E-28040, Spain\\ 
$^2$ Racah Institute of Physics, The Hebrew University, Jerusalem, 91904, Israel\\ 
$^3$ University of California, Santa Cruz, CA 95064, USA\\ 
$^4$ Universit{\"a}t Heidelberg, Zentrum f{\"u}r Astronomie, Institut f{\"u}r Theoretische Astrophysik, Albert-Ueberle-Str. 2, 69120, Heidelberg,\\ Germany\\ 
$^5$ New York University, Abu Dhabi, PO Box 129188, Saadiyat Island, Abu Dhabi, United Arab Emirates\\ 
$^6$ Max-Planck-Institut f{\"u}r Astronomie, K{\"o}nigstuhl 17, D-69117, Heidelberg, Germany\\ 
$^{7}$ Department of Astronomy and Astrophysics, UCO/Lick Ob- servatory, University of California, 1156 High Street, Santa Cruz,\\ CA 95064, USA\\ 
$^{8}$ Kavli Institute for the Physics and Mathematics of the Universe, University of Tokyo, Kashiwa, Chiba 277-8583, Japan\\ 
$^{9}$ CIERA Fellow, Department of Physics and Astronomy and CIERA, Northwestern University, Evanston, IL, USA
}
\date{Accepted XXX. Received 10/2018; in original form 08/2018}
\pubyear{2018}
\begin{document}
\label{firstpage}
\pagerange{\pageref{firstpage}--\pageref{lastpage}}
\maketitle

\begin{abstract}
We study the components of cool and warm/hot gas in the circumgalactic medium (CGM)  of simulated galaxies and address the relative production of OVI by photoionization versus collisional ionization, as a function of halo mass, redshift, and distance from the galaxy halo center.  This is done utilizing two different suites of zoom-in hydro-cosmological simulations, VELA (6 halos; $z>1$) and NIHAO (18 halos; to $z=0$), which provide a broad theoretical basis because they use different codes and physical recipes for star formation and feedback.
In all halos studied in this work, we find that collisional ionization by thermal electrons dominates at high redshift, while photoionization of cool or warm gas by the metagalactic radiation takes over near $z\sim2$. In halos of $\sim 10^{12}M_{\odot}$ and above, collisions become important again at $z<0.5$, while photoionization remains significant down to $z=0$ for less massive halos. In halos with $M_{\textrm v}>3\times10^{11}~M_{\odot}$, at $z\sim 0$ most of the photoionized OVI is in a warm, not cool, gas phase ($T\lesssim 3\times 10^5$~K). We also find that collisions are dominant in the central regions of halos, while photoionization is more significant at the outskirts, around $R_{\textrm v}$, even in massive halos. This too may be explained by the presence of warm gas or, in lower mass halos, by cool gas inflows.
\end{abstract}




\section{Introduction}\label{sec:1}
Gas in the circumgalactic medium (CGM) is a key ingredient in galaxy evolution. Gas from the intergalactic medium (IGM) streams into the centers of DM halos and feeds star formation. Stellar feedback heats up cold gas and produces metallicity-enhanced warm/hot gas ($T>10^{4.5}$~K) that outflows to the CGM. Theory and simulations agree that halos with $M>M_{\textrm crit}\sim 3 \times 10^{11}~M_{\odot}$ develop a warm/hot CGM through both shock heating at $\sim R_{\textrm v}$ and outflows \citep{DekelBirnboim06}. In these systems cold gas in narrow inflowing streams can penetrate through warm/hot CGM only at $z>2$. In low mass halos ($M<M_{\textrm crit}$), however, the CGM is always dominated by cold  ($T~<10^{3.8}$~K) and cool gas (10$^{3.8}$~K~$<T<10^{4.5}$~K) \citep{BirnboimDekel03,Keres05,DekelBirnboim06,Keres09,Correa17,Fielding17}. This implies that halos of mass $\sim10^{12}$ and above should have a significant fraction of their baryons in a warm-hot CGM, which may serve as a gas reservoir for star formation at later times \citep{Faerman17}. The understanding of the formation, evolution and properties of the CGM is thus mandatory for the study of the evolution of galaxies. \smallskip \\

Even if massive, this warm/hot gas component is difficult to detect as its density is low and its emission diffuse. In addition, most of the radiation is at high energies, in the UV and X-ray ranges, which are hard to observe \citep[e.g.,][]{Crain10}. The most common technique used to study the diffuse CGM is to look for absorption lines in background bright sources (e.g., QSO). This method has been useful for the detection of multiphase (cool and hot) gas in the CGM of galaxies, but it requires a close projection between the studied galaxy and the background source. Low temperature and high density gas has been observed through lines from gas in lower ionization states or from Lyman Limit Systems (LLS) and Damped Lyman-Alpha systems (DLAs). Warm/hot diffuse gas can only be studied by observations of lines from highly ionized metals. Each observation gives only a single measurement of CGM properties thus a large number of observations are required to get a general picture of the CGM surrounding galaxies. Many useful ion absorption lines probing CGM temperatures fall in the UV range; as a consequence, it is easier to study properties of the CGM in high-redshift galaxies where the rest-frame UV absorption lines have been redshifted to the optical range \citep[e.g.,][]{CowieSongalia95,Steidel10,Rudie13,Lehner14}. This is the case for OVI lines that for low redshift galaxies ($z<2.0$) fall in the UV spectral range.  The most relevant observations in low-redshift CGMs have utilized the Cosmic Origins Spectrograph (COS) on the Hubble Space Telescope (HST) \citep[e.g.,][]{ThomChen08,Prochaska11}. In particular, the COS-Halos project provides measurements of highly ionized atoms, including OVI, for the CGMs of a large sample of $\sim L_*$ galaxies \citep{Tumlinson13}.\smallskip \\

The interpretation of the composition of the CGM based on a single metal ion is not straightforward. The determination of the distribution of mass, temperature, and density of the gas requires understanding of the ionization mechanism at work. Oxygen is collisionally ionized (CI, hereafter) to OVI between 10$^5$~K~$<T<10^6$~K, nearly independent of density. On the other hand, at lower densities and lower temperatures it may be produced by photoionization (PI, hereafter), once a hard external radiation field is present. Thus, OVI in the CGM can be a tracer of either collisionally-ionized warm/hot gas or photo-ionized cool gas \citep{Savage02,Peeples14,Werk14,Stern16,Faerman17}. The former could have originated, for example, from feedback-driven galactic outflows \citep{Mathews17} or from virial shock heating, and the latter could arise, e.g., from cold gas inflows \citep{Stern18} or from cooling of the warm-hot phase \citep{Bordoloi17}. Observers dispute which is the gas phase that dominates the CGM. On one hand, X-ray observations of higly ionized oxygen emission lines \citep{Henley10,HenleyShelton10} and also absorption lines within the spectrum of extremely bright QSOs \citep[e.g][]{Nicastro02,Rasmussen03}, suggest the presence of extended massive warm/hot coronae dominated by CI high ions (e.g. OVII/OVIII). Furthermore, extended X-ray line and continuum emission observed in low-$z$ galaxies \citep{AndersonBregman11,Bogdan13,Anderson15} points towards that direction. On the other hand, a detailed examination of the CGM absorption line profiles \citep{Tripp08,ThomChen08,Werk14,Stocke13} have shown a high covering factor of photoionized gas clouds in the CGM of low-$z$  star-forming galaxies. Also, COS-Halos found that it is possible that a large fraction of the CGM gas mass is in a cool low ionization state. This cool gas appears to be in dense knots that are not in hydrostatic equilibrium with their surroundings \citep{Werk14}. How such gas is supported is not well understood. A combined scenario of both a warm/hot and a cool CGM has been also proposed by \citet{Stocke14} and \citet{Pachat16}, who postulated that cool PI clouds can be embedded in a hotter more massive CI diffuse gas, based on OVI and Ly$\alpha$ absorption line detections. \smallskip \\

Hydrodynamical simulations have been used to study the CGM properties, the OVI distribution, and to attempt to break the degeneracy concerning its production mechanism. Some simulations obtained an OVI column density (N$_{\textrm OVI}$, hereafter) that is lower by an order of magnitude from the observed value \citep[e.g.,][]{Hummels13,Suresh15,RocaFabrega16,Ford16,Oppenheimer16}. Other simulations reveal better, although still not full, agreement with the observations \citep[e.g.,][]{Gutcke17,Suresh17,Nelson18}. Some of these works also found a clear dependence of the N$_{\textrm OVI}$ on the star formation rate and the $M_{200}$ of the host galaxy, and on its environment \citep{Oppenheimer16, Liang16}. Feedback efficiency showed to play a key role in modifying the shape and normalization of the N$_{\textrm OVI}$ profile, due to its effect on the metal production \citep{Rahmati16,Oppenheimer16}. Also it has been found that the total OVI mass and distribution depends on the feedback strength \citep[e.g. ][]{Liang16}. Feedback strength determines the amount of warm/hot gas that is produced, and how much that generates OVI through CI \citep{Fielding17}. Simulations have also indicated that in low-mass halos the OVI is produced by PI more than in high-mass halos \citep[e.g.][, at $z=0$]{Gutcke17}. Recent works also agree on that in halos with masses larger than 10$^{12}M_{\odot}$, at low-$z$, OVI is mainly produced through collisional ionization \citep{Liang18}. These results agree with the theoretical framework proposed by \citet{BirnboimDekel03} where higher mass halos are able to generate a warm/hot CGM through virial shock heating while lower mass systems are dominated by cold/cool flows. \smallskip \\
In this work we use two simulation suites to study the CGM as a function of redshift, halo mass, and distance from the halo center. In Sec.~\ref{sec:2} we describe the simulation suites. In Sec.~\ref{sec:4} and ~\ref{sec:4.7} we analyze properties of the CGM gas in simulations and we study the OVI distribution as function of halo mass, radius and redshift. The summary and conclusions are presented in Sec.~\ref{sec:8}. \smallskip \\

\section{Simulations}\label{sec:2}
We use two suites of simulations, VELA and NIHAO. VELA is run using the ART code  \citep{Kravtsov97,Kravtsov03} and NIHAO using GASOLINE2 \citep{Wadsley17}. Both of them obtain high resolution by zooming in on one halo at a time inside a fully cosmological box. In the following sections we describe the main properties of these simulations. 

\subsection{VELA}\label{sec:2.1}
The first set of simulations we used is a subsample of 6 galaxies from the VELA suite. The entire VELA suite contains 35 halos with virial masses ($M_{\textrm v}$)\footnote{All virial quantities we show in this paper are computed assuming $R_{\textrm v}=R_{\textrm {200}}$, where $R_{\textrm {200}}$ is the radius where halo density is 200 times the critical density of the Universe.} between $2\times10^{11}~M_{\odot}$ and $2\times10^{12}~M_{\odot}$ at $z=1$. The VELA suite was obtained using the ART code \citep{Kravtsov97,Kravtsov03}, which follows the evolution of a gravitating N-body system and the Eulerian gas dynamics using an AMR approach. Beyond gravity and hydrodynamics, the code incorporates many of the physical processes relevant for galaxy formation. These processes, representing sub-grid physics, include gas cooling by atomic hydrogen, helium, metals, and molecular hydrogen, photoionization heating by a constant cosmological UV-background with partial self-shielding (self-shielding is applied when gas reaches a density of 0.1~cm$^{-3}$), stochastic star formation and SN feedback, as described in \citep{CeverinoKlypin09,Ceverino10,Ceverino14}. Star formation occurs only in regions with density above 1~cm$^{-3}$ and $T<10^4$~K. In addition to thermal-energy supernova feedback the simulations incorporate radiative feedback. This model adds a non-thermal pressure, radiation pressure, to the total gas pressure in regions where ionizing photons from massive stars are produced and trapped. In these simulations the dark matter particle minimum mass is of $8.3\times10^4~M_{\odot}$, while the one of star particles is $10^3~M_{\odot}$. The maximum spatial resolution is between 17$-$35 physical pc. All details about the VELA suite can be found in \citet{Ceverino14,Zolotov15}. \smallskip \\
We selected our subsample according to its total virial mass and the final redshift the simulation reached. Here we use all halos that have been simulated down to $z=1$ and that have a final mass $>5\times10^{11}~M_{\odot}$. This selection criteria derives from our interest in studying the variation of CGM properties both as function of redshift and virial mass around $M_{\textrm {crit}}$. Once the halos become more massive than several times $10^{11}~M_{\odot}$ they show a transition from the cold flow regime to the shock-heated regime, i.e. a warm/hot gas corona is generated \citep{GoerdtCeverino15,Zolotov15}. Lower mass halos are dominated by cold flows at all-$z$  as they do not generate virial shocks \citep{BirnboimDekel03}. Properties of the six VELA runs used in this work are described in Table ~\ref{tab:1}. \smallskip \\

\begin{table}
\begin{centering}
 \begin{tabular}{l c c c c} 
 \hline\hline
VELA  & $M_{\textrm v}$ & $M_*$ & $M_{\textrm g}$ & $R_{\textrm v}$ \\
  & [$10^{12}M_{\odot}$] & [$10^{10}M_{\odot}$] & [$10^{10}M_{\odot}$] & [kpc]  \\
 \hline
 {\bf V07} & {\bf 1.45} & {\bf 13.3} & {\bf 6.31} & {\bf 164}  \\ 
 V08 & 1.12 & 6.03 & 5.96 & 151 \\
 V10 & 0.64 & 3.01 & 4.09 & 125 \\
 V21 & 0.83 & 7.12 & 2.81 & 137 \\
 V22 & 0.61 & 5.13 & 1.16 & 123 \\
 V29 & 0.70 & 3.56 & 3.04 & 128  \\
 \hline\hline
\end{tabular}
\label{tab:1}
 \caption{Properties at $z=1$ of the VELA simulations used in this work. The virial quantities are computed at $R_{\textrm {200}}$.}
 \end{centering}
\end{table}

\subsection{NIHAO}\label{sec:2.2}
The second set of simulations we use in this work is a subsample of 18 galaxies from the NIHAO suite of simulations, consisting of 88 halos with a large mass range \citep{Wang15}. The NIHAO simulations have been obtained using a revised version of the N-body SPH code GASOLINE, named ESF-gasoline2 \citep{Wang15, Wadsley17}. In these simulations the dark matter particle minimum mass ranges from $2.1\times10^5$ to $1.7\times10^6M_{\odot}$, depending on the zoom-in model (see \citet{Wang15} for a detailed discussion about spatial and mass resolution). The sub-grid physics recipes used include diffusion of metals as described in \citet{Wadsley08} and cooling via hydrogen, helium, and various metal lines in a uniform ultraviolet ionizing background as described in \citet{Shen10} and calculated using Cloudy \citep[version 07.02][]{Ferland98}. No shielding is imposed. Star formation occurs when the gas temperature and density reach an imposed threshold. Fiducial thresholds used in NIHAO simulations are $T<1.5\times 10^4$~K and $n>10.3$~cm$^{-3}$. The supernova feedback follows the blast-wave formalism \citep{Stinson06}. To avoid overcooling, cooling is delayed for particles inside the blast region for $\sim$30~Myr. Also early stellar feedback by massive stars is used \citep{Stinson13}. A full description of sub-grid physics and the success of NIHAO simulations in reproducing observed galaxy properties can be found in \citet{Wang15} and \citet{Dutton17} (NIHAO XII). As a general result, star formation parameters and SN recipes used in NIHAO result in stronger feedback than VELA. \smallskip \\
The selection criteria we used to select our subsample is the same as the one for VELA simulations. We have selected halos with a virial mass $>5\times 10^{11}~M_{\odot}$ at $z=1$. Unlike the VELA suite, all NIHAO runs reach $z=0$. We also selected two halos with masses below $M_{\textrm crit}$ at $z=0$, one with $M_{\textrm v}\sim10^{11}~M_{\odot}$ and another with $M_{\textrm v}\sim 10^{10}~M_{\odot}$. These models never develop a warm/hot CGM and they are used to be compared with results from more massive galaxies. The final subsample we used in this work contains 18 halos. Properties of the 18 halos at $z=1$ and $z=0$ are described in Table~\ref{tab:2}. \smallskip \\

\begin{table*}
\begin{centering}
 \begin{tabular}{l r r r c | r r r c} 
 \hline\hline
NIHAO & \multicolumn{1}{c}{$M_{\textrm v}$} & \multicolumn{1}{c}{$M_*$} & \multicolumn{1}{c}{$M_{\textrm g}$} & \multicolumn{1}{c|}{$R_{\textrm v}$} & \multicolumn{1}{c}{$M_{\textrm v}$} & \multicolumn{1}{c}{$M_*$} & \multicolumn{1}{c}{$M_{\textrm g}$} & \multicolumn{1}{c}{$R_{\textrm v}$} \\
  & \multicolumn{1}{c}{[10$^{12}M_{\odot}$]} & \multicolumn{1}{c}{[10$^{10}M_{\odot}$]} & \multicolumn{1}{c}{[10$^{10}M_{\odot}$]} & \multicolumn{1}{c|}{[kpc]} & \multicolumn{1}{c}{[10$^{12}M_{\odot}$]} & \multicolumn{1}{c}{[10$^{10}M_{\odot}$]} & \multicolumn{1}{c}{[10$^{10}M_{\odot}$]} & \multicolumn{1}{c}{[kpc]}\\
  & \multicolumn{1}{c}{(z=1)} & & & & \multicolumn{1}{c}{(z=0)} & & & \\
 \hline
 g2.79e12 & 2.24 & 9.44 & 18.52 & 203     & 4.42 & 20.12 & 34.69 & 510 \\ 
 {\bf g1.92e12 (N01) }&  {\bf 1}. {\bf 85 }& {\bf 10}.{\bf 00} & {\bf 13}.{\bf 81 }& {\bf 191  }   & {\bf 2}.{\bf 90 }& {\bf 15}.{\bf 91 }& {\bf 18}.{\bf 37 }& {\bf 444} \\
 g1.77e12 & 0.91 & 1.72 & 10.35 &  151    & 2.68 & 13.84 & 18.98 & 432 \\
 g1.12e12 & 0.82 & 3.90 & 6.10 & 145    & 1.50 & 7.95 & 10.77 & 356 \\
 g8.26e11 & 0.77 & 3.36 & 6.09 & 142     & 1.27 & 4.74 & 10.23 & 337 \\
 g8.13e11 & 0.81 & 4.41 & 5.69 & 145     & 1.22 & 6.73 & 7.32 & 333 \\
 g8.06e11 & 0.45 & 0.45 & 4.55 & 119     & 1.20 & 4.51 & 9.87 & 330 \\
 g7.66e11 & 0.70 & 0.91 & 8.02 & 138     & 1.12 & 5.96 & 6.47 & 323 \\
 g7.55e11 & 0.70 & 1.05 & 7.51 & 138     & 1.20 & 3.16 & 11.30 & 330 \\
 g7.44e11 & 0.28 & 0.18 & 2.99 & 102     & 1.62 & 1.95 & 18.18 & 365 \\
 g7.08e11 & 0.46 & 1.43 & 4.20 & 120     & 1.09 & 3.14 & 9.30 & 320 \\
 g6.96e11 & 0.49 & 0.24 & 5.80 & 122     & 1.10 & 3.40 & 8.16 & 321 \\
 g5.55e11 & 0.33 & 0.16 & 3.71 & 108     & 0.68 & 1.76 & 4.29 & 273 \\
 g5.38e11 & 0.52 & 0.83 & 4.71 & 125     & 0.82 & 1.87 & 5.66 & 292 \\
 g5.36e11 & 0.20 & 0.14 & 2.06 & 91     & 0.97 & 1.23 & 8.99 & 308 \\
 g5.31e11 & 0.41 & 0.41 & 4.10 & 115     & 0.72 & 1.66 & 6.36 & 278 \\
 g5.02e11 & 0.45 & 0.45 & 4.54 & 119     & 0.76 & 1.48 & 5.90 & 283 \\
  \hline
 g2.19e11 & 0.07 & 0.02 & 0.71 & 64     & 0.17 & 0.09 & 1.55 & 172 \\ 
 g2.63e10 & 0.023 & 0.0037 & 0.071 & 44     & 0.034 & 0.0043 & 0.066 & 100 \\ 
 \hline\hline
\end{tabular}
\label{tab:2}
 \caption{Properties at $z=1$ and $z=0$ of the NIHAO simulations used in this work. Virial quantities are computed at $R_{\textrm {200}}$.}
 \end{centering}
\end{table*}

\section{CGM gas properties in simulations}\label{sec:4}
In this section we analyze the spatial distribution and properties of CGM gas in the VELA and NIHAO simulations. We study the distribution in mass, volume, temperature, velocity, and metallicity space. \smallskip \\

\subsection{Definitions and initial considerations}\label{sec:4.1}
Most figures in this work show stacked results for each simulation suite. When stacking is not possible, we show results from two representative simulated galaxies, VELA07 (V07) and NIHAO g1.92e12 (N01), which have similar masses throughout their evolution from $z=4$ to $z=1$. We address evolution with redshift by showing snapshots at $z=2.3$ and $z=1$ for the two simulation suites, as well as at $z=0$ for NIHAO.  \smallskip \\
In accordance with previous works \citep{Stern16,Faerman17}, we distinguish between the following three gas phases by temperature: \smallskip \\
\begin{itemize}
\item Cold gas: $T < 10^{3.8}$~K
\item Cool gas: 10$^{3.8}$~K~$<T<10^{4.5}$~K
\item Warm/hot gas: 10$^{4.5}$~K~$<T<10^{6.5}$~K
\end{itemize}

\begin{figure*}
   \centering
   \includegraphics[scale=0.3]{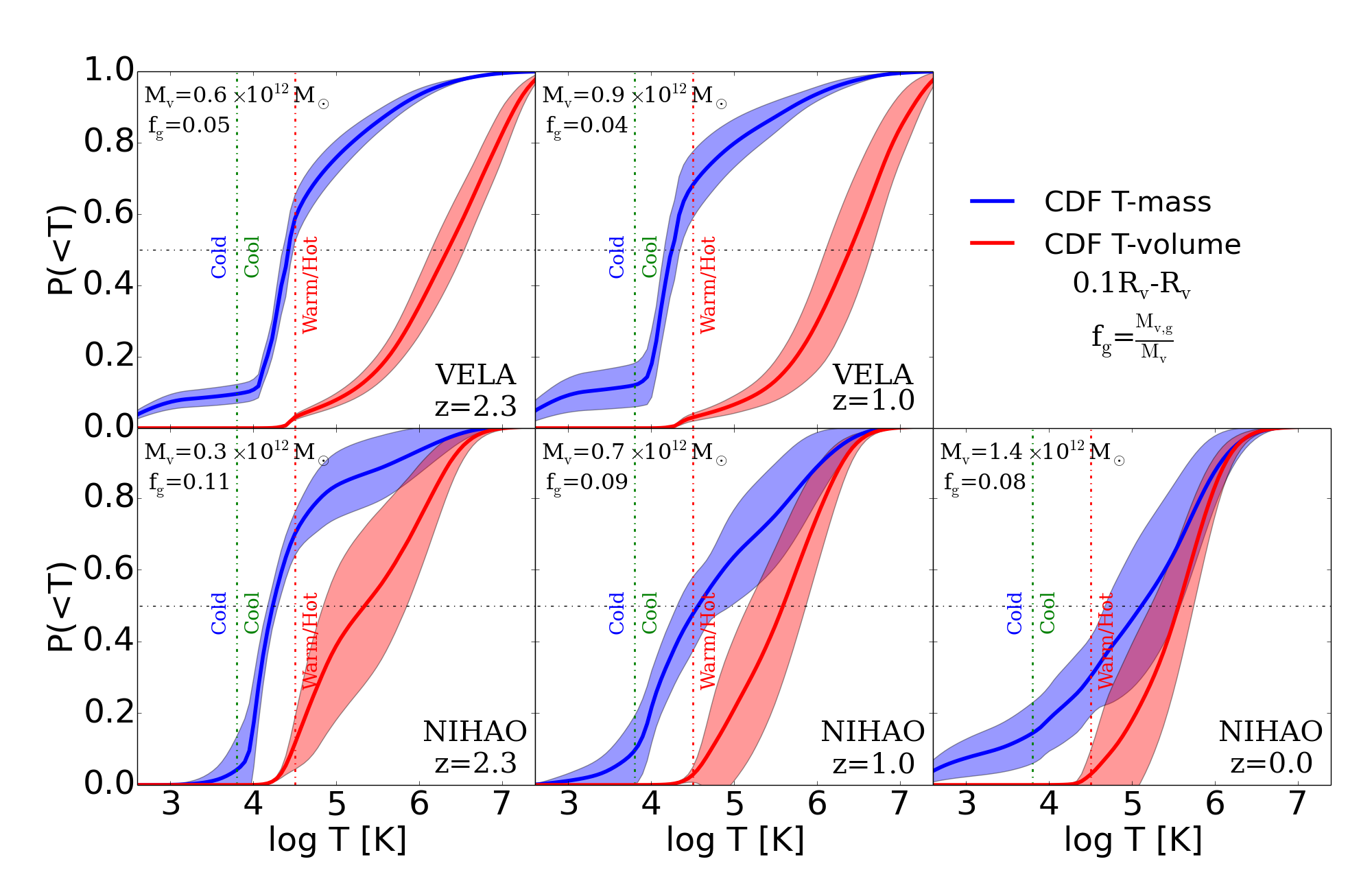}
   \caption{Mean mass/volume (blue/red) cumulative distribution function (CDF) as function of temperature, for VELA (top row) and NIHAO (bottom row), stacked over all halos. Each column shows results at a different redshift (i.e., $z=2.3$, 1.0 and 0.0, from left to right). CDFs have been computed using all gas inside $0.1R_{\textrm v}-R_{\textrm v}$. Mean virial mass ($M_{\textrm {v}}$) and gas fraction ($f_{\textrm g}=M_{\textrm {v,g}}/M_{\textrm {v}}$) are indicated in top-left corner of each panel. Solid lines show the mean CDF. The shadowed region shows the 1$\sigma$ dispersion from the mean. Thin dot-dashed vertical lines indicate the transition temperature between different gas phases, cold to cool in green and cool to warm/hot in red. Each phase is labeled following the same color convention. The thin horizontal dot-dashed line is just a guiding line at 50\%.}
   \label{fig:1}%
\end{figure*}

\subsection{Temperature-density distribution}\label{sec:4.2}
In this section we focus on the distribution of temperature and density in the CGM gas and its evolution with $z$. \smallskip \\

\subsubsection{The temperature cumulative distribution functions (CDF)}\label{sec:4.2.1}
In Fig.~\ref{fig:1} we show the mass-weighted (blue) and volume-weighted (red) mean cumulative gas distribution functions (CDF) of temperature in the CGM (0.1$-1.0R_{\textrm v}$). The top row of Fig.~\ref{fig:1} shows the CDF for the stacked VELA halos at $z=2.3$ (left) and 1.0 (right). The bottom row shows the stacked CDF for NIHAO at $z=2.3$ (left), 1.0 (middle) and 0.0 (right). Solid lines show mean CDF values and shadowed regions the 1$\sigma$ dispersion. Vertical lines indicate the transition between gas phases. \smallskip \\
We first focus on the mass-weighted CDF (blue). In the mass-CDF plot three gas phases are present i.e., cold, cool and warm/hot. These phases are better differentiated at higher redshift and in VELA. In simulations presented here the cold component (with $T<10^4$~K) never reaches values above 10\% of the total mass. Cold and cool gas ($T<10^{4.5}$~K) is the largest component by mass; at $z=2.3$ it contains between 50 and 70\% of the total gas mass, both in NIHAO and VELA. Warm/hot gas with temperatures above 10$^{4.5}$~K contains a small fraction of the total mass at high-$z$ and eventually includes about 50$-$80\% at mid-low-$z$. The difference between the total warm/hot gas in VELA and NIHAO may suggest a dependence on feedback strength \citep{BirnboimDekel03,Keres05,Ceverino14}. \smallskip \\
The volume-weighted temperature distribution shows only two well defined gas phases: a cold/cool phase that fills a tiny fraction of the total volume and a warm/hot phase that fills almost the entire volume. This result suggests that the cold/cool gas is mostly in and around compact dense regions while the warm/hot gas fills the entire CGM volume. This is true for the two suites of simulations and is almost independent of redshift. \smallskip \\
As previously mentioned, the CGM gas mass distribution versus temperature in both the VELA and NIHAO runs changes with time. This evolution is driven by an increase of warm/hot gas mass in massive halos at low-$z$. In all simulations the cool gas mass remains constant or decreases slightly with decreasing $z$. \smallskip \\
We note certain differences between VELA and NIHAO halos. First, the VELA halos have a cold gas component of 10\% in mass, already at $z=2.3$, while the NIHAO halos develop such a component only at lower redshifts. Second, gas is depleted, i.e. consumed through star formation or expelled to the IGM/CGM by SNe feedback, more efficiently in VELA than in NIHAO. This makes the gas fraction to be two times larger in NIHAO than in VELA. Both differences can be attributed to: the UVB shelf-shielding of dense, cold gas in VELA, to the presence of early stellar feedback by massive stars in NIHAO, or to a combination of both. The global consequence of such differences is that in NIHAO gas cools into the star forming regions less efficiently, as it is heated up by the UVB. \smallskip \\

\begin{figure*}
   \centering
   \includegraphics[scale=0.24]{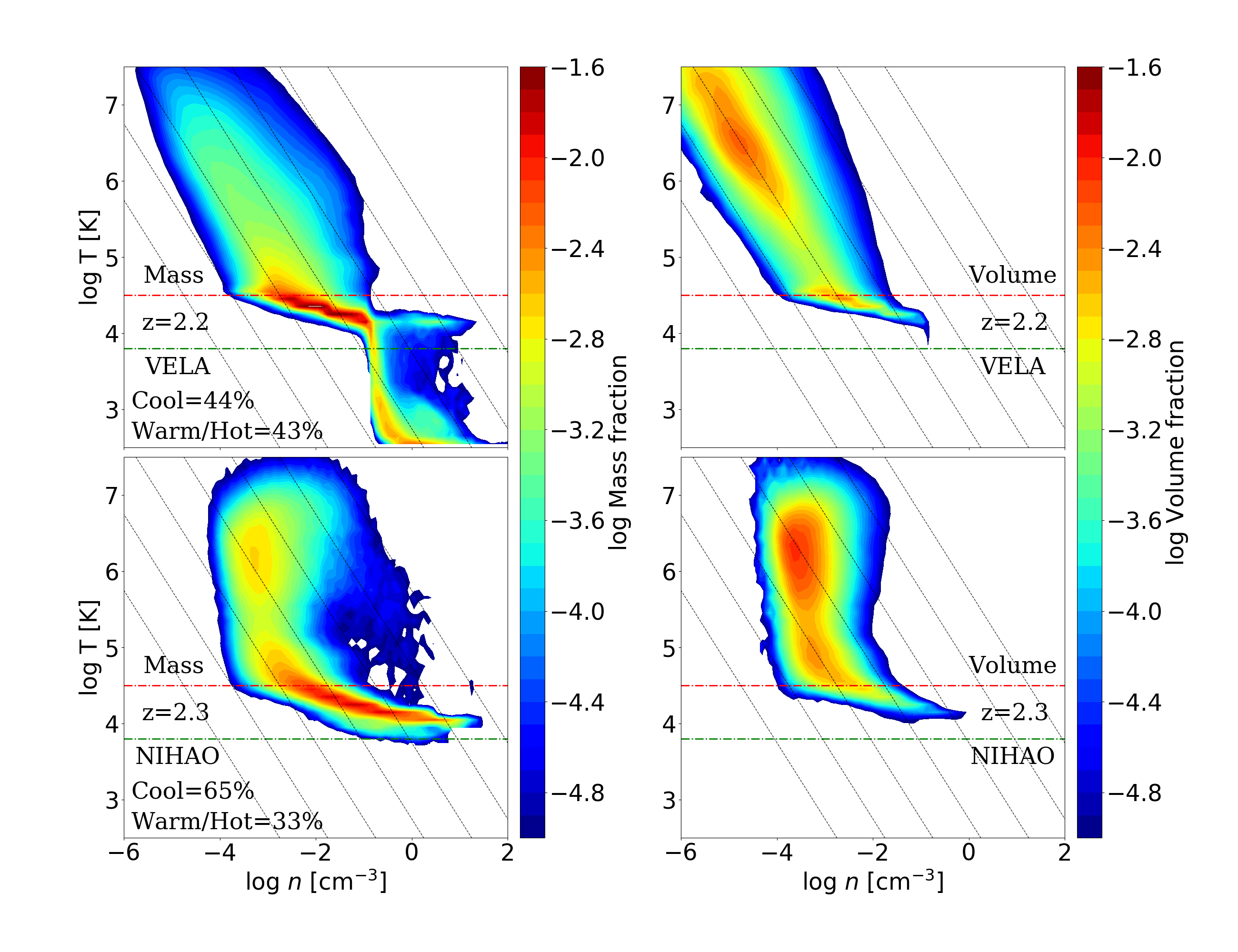}
   \caption{$T$ vs. $\rho$ diagrams of the total gas mass (left) and volume (right), inside 0.1$-1.0~R_{\textrm v}$. Top row shows results after stacking the VELA runs at $z=2.2$. Bottom row shows the same as top but for the NIHAO runs at $z=2.3$. Colors show the mass/volume fraction inside a temperature-density cell of 0.1~$\times$~0.1~dex. Diagonal gray dashed lines show isobaric evolution. Horizontal lines show transitions between gas phases: cold to cool in green and cool to warm/hot in red. In this figure differences between the two simulation techniques are apparent (e.g., a density limit at $log n=$-1 on the UV-background heating is imposed in VELA to account for self-shielding but not in NIHAO, or, a delayed cooling gas phase is present in NIHAO at $T>5\times 10^4$~K and $log n>-1.5$, but not in VELA.}
   \label{fig:2}%
\end{figure*}

\subsubsection{Phase-space diagrams $T-\rho$}\label{sec:4.2.2}
Fig.~\ref{fig:2} shows stacked $T-\rho$ diagrams weighted by mass (left) or volume (right). We show the total gas mass per temperature-density bin (left) and volume (right) at $z\sim~2.2$ and for VELA (top) and NIHAO (bottom), inside 0.1$-1.0~R_{\textrm v}$ (CGM region). Gas in the disk region (gas at $R<0.1~R_{\textrm v}$, that is not included in Fig.~\ref{fig:2}) has, in NIHAO models, two components, i.e. hot-dense and cold-dense. Hot gas component is produced by heating from supernovae and kept dense due to the delayed cooling regime \citep{Stinson15}. Cold-dense gas is gas cooled down by implemented cooling mechanisms. In VELA this cold-dense gas component is also present in the disk region and is more massive than in NIHAO. Differences in the cold-dense gas component in NIHAO and VELA derive from variations in the self-shielding recipe as discussed in Sec.~\ref{sec:4.2.1}. \smallskip \\
The cold and cool CGM gas components contain most of its mass at high and mid redshift, both in VELA and NIHAO, as seen in the temperature CDF (Fig.~\ref{fig:1}). The cool gas mass is distributed over a wide range of densities and it fills a small fraction of the volume. The cool gas seems to be located at the ends of isobaric cooling paths (grey dashed lines) of warmer gas with a wide range of densities. The gas cools efficiently down to $T\sim 10^4$~K, where the cooling curve drops. As the gas accumulates at $T\sim 10^4$~K and its density grows, self-shielding is turned on (in VELA) and slow metal cooling continues down to the cold phase (see further discussion in the following sections). The mass fraction in the warm/hot phase grows in time, more so in NIHAO. Most of the CGM volume is filled by this warm/hot gas component both in NIHAO and VELA. In almost all the VELA halos the distribution of warm/hot gas seems to follow isobaric lines. This result suggests that cooling from hot to warm gas probably occurs at constant pressure. The non-trivial presence of CGM warm gas on the peak of the cooling curve ($T \sim5\times10^4$~K), i.e. with short lifetimes, in all simulations presented here, will be discussed below. \smallskip \\

\subsection{Spatial distribution}\label{sec:4.3}

Fig.~\ref{fig:5_2} and \ref{fig:5_3} display the spatial distribution of cool and warm/hot gas in V07 (top 4 panels) and N01 (bottom panels) at $z=1.0$. In the top-left panel of each set of four we show the corresponding $T-\rho$ diagram. In the top-right and bottom panels we show the projected density in three orthogonal directions. \smallskip \\

\subsubsection{Cool gas}

Fig.~\ref{fig:5_2} presents the cool gas distribution in V07 (top) and N01 (bottom), at $z=1.0$, inside $R_v$. After analyzing the temporal evolution of gas in the real space, both inside and outside $R_v$, we conclude that the cool gas is mostly in filaments and flowing down to the disk. In V07 the filaments are broad and fill a large fraction of the CGM while in N01 the cool gas is in thin and clumpy filaments, filling a small fraction of the CGM. Filaments in N01 appear broader if we include gas at slightly higher temperatures. A small fraction of cool gas in VELA is also in outflows (see \citep{Ceverino16b} for a full discussion of cool gas outflows).\smallskip \\

\subsubsection{Warm/Hot gas}

\begin{figure}
   \centering
   \includegraphics[scale=0.30]{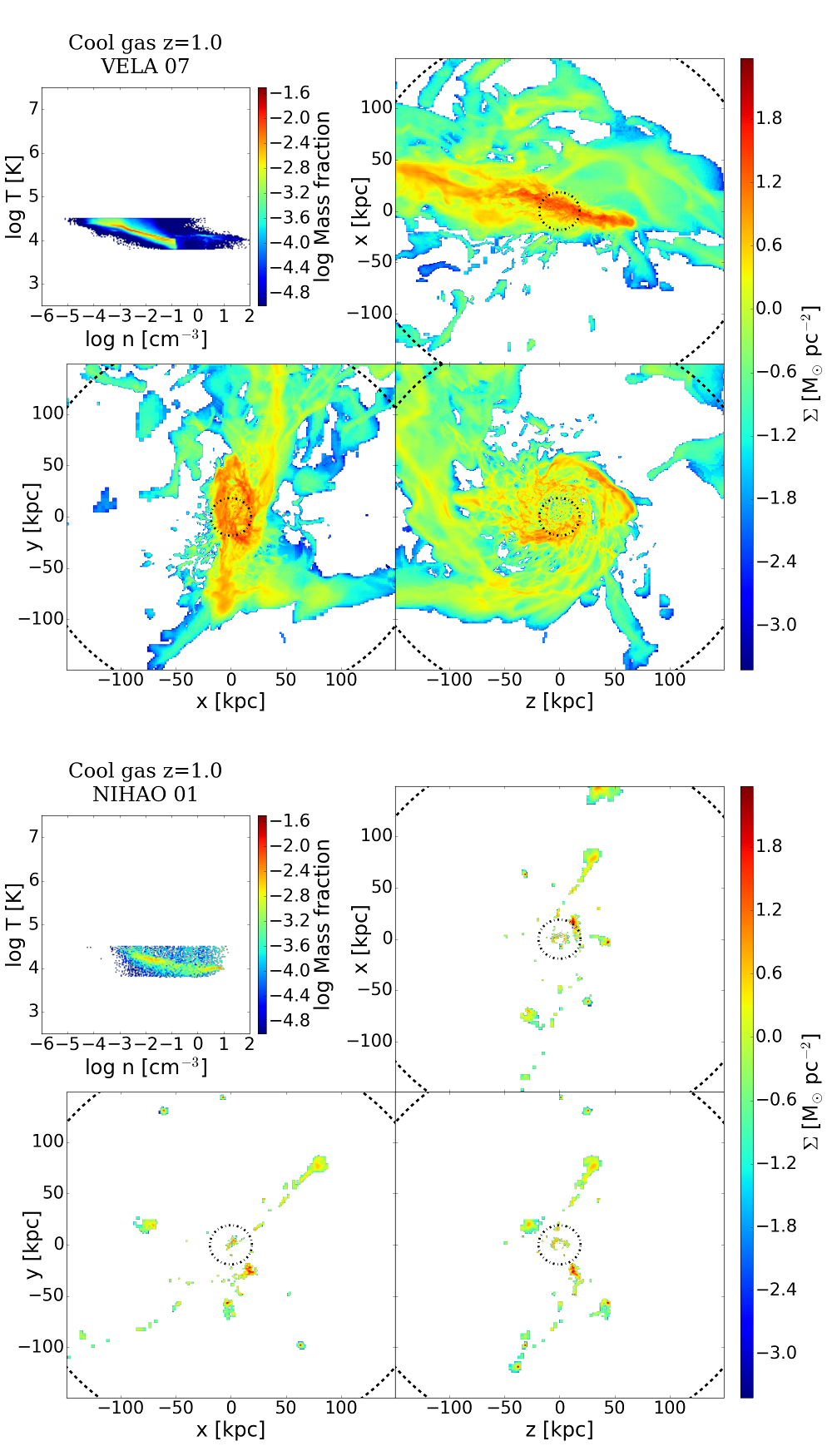}
   \caption{Cool gas mass distribution in V07 (top four panels) and N01 (bottom panels) at $z=1$. In each set of 4 panels we show three orthogonal density projections in the coordinates space of all gas within 1.5~$R_{\textrm v}$ (top-right and bottom panels) and the corresponding $T-\rho$ diagram (top-left panel). Colors show projected gas density/gas mass. Dotted circles indicate 0.1~$R_{\textrm v}$ and $R_{\textrm v}$, respectively. The two halos shown here have been selected because they have similar mass ($M_v$ and $M_*$) and as they are representative of the general behavior among the entire set of models.}
   \label{fig:5_2}%
\end{figure}

\begin{figure}
   \centering
   \includegraphics[scale=0.30]{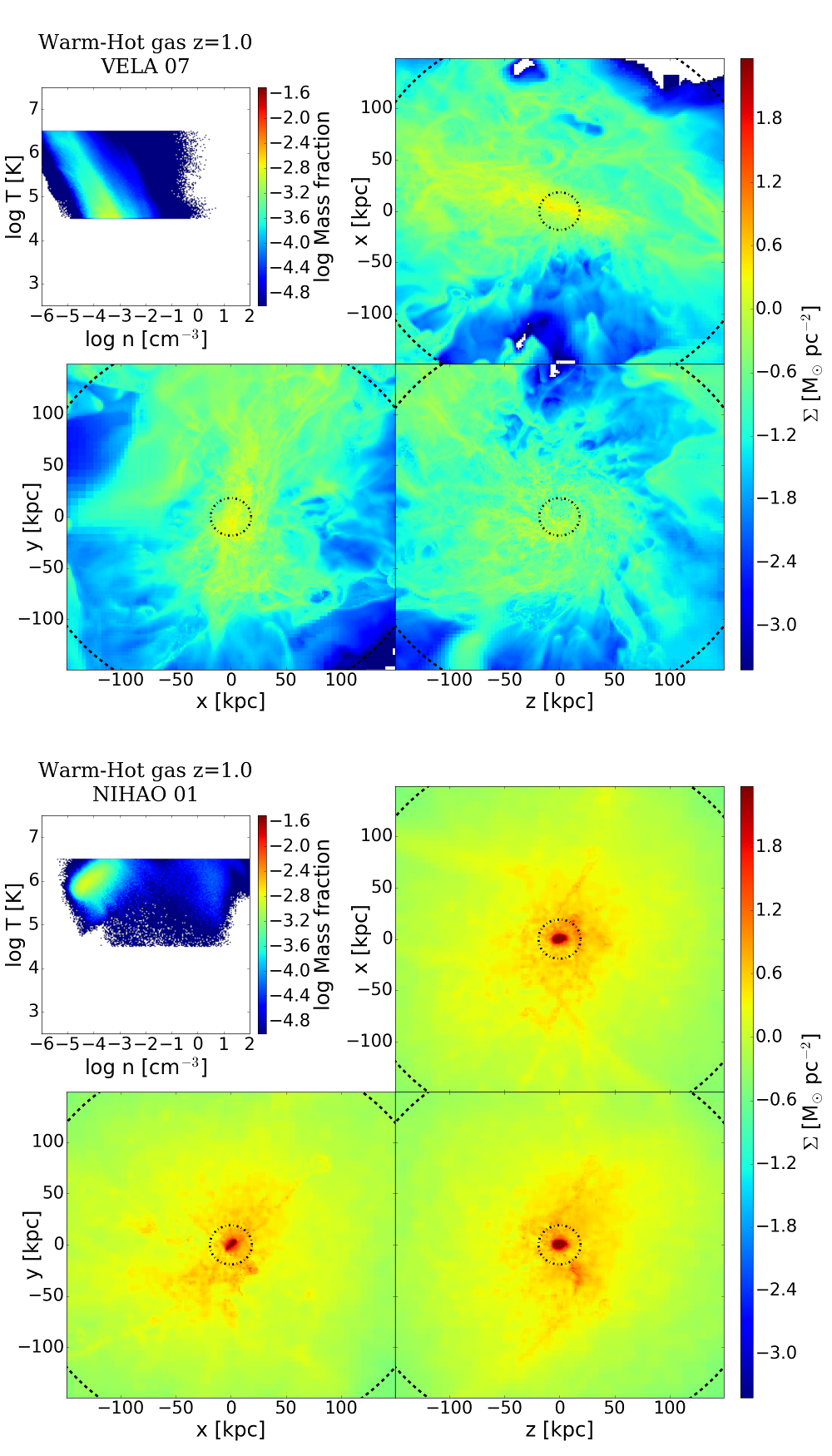}
      \caption{Same as Fig.~\ref{fig:5_2} but for warm/hot gas phase.}
   \label{fig:5_3}%
\end{figure}

Fig.~\ref{fig:5_3} shows the same as Fig.~\ref{fig:5_2} but for the warm/hot gas component which fills the CGM volume in a much more  homogeneous distribution than the cool gas. In NIHAO, we see the central hot dense component, as indicated in the phase-space diagram (Fig.~\ref{fig:2}), resulting from the delayed cooling of the supernova remnant. This gas rapidly cools down when reaching larger distances in the CGM. As indicated in the phase-space diagrams, NIHAO shows a denser warm/hot CGM than VELA. This arises from the stronger feedback and possibly less efficient cooling in NIHAO. \smallskip \\


\subsection{Metallicity distribution}\label{sec:4.4}

\begin{figure*}
   \centering
   \includegraphics[scale=0.24]{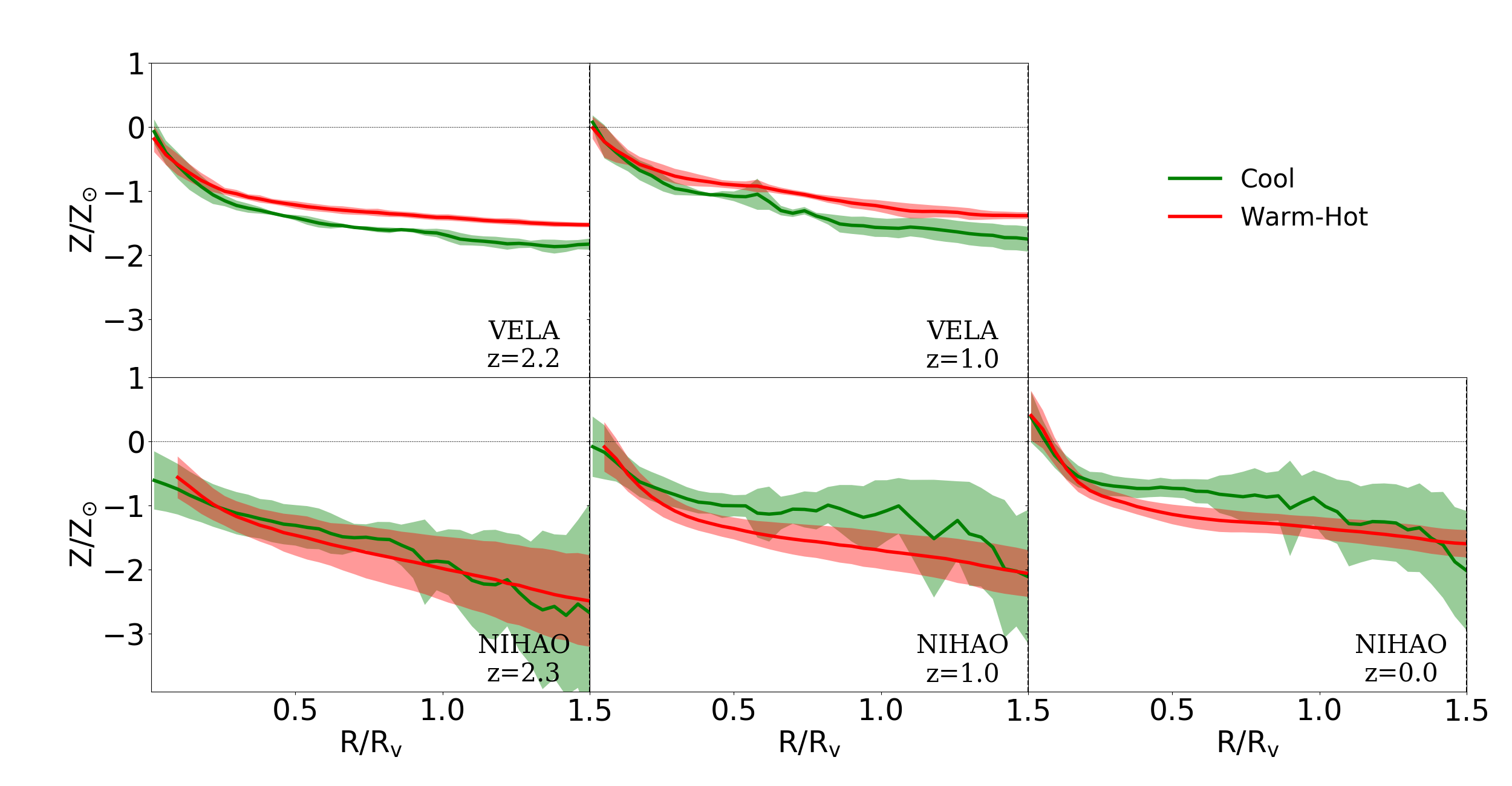}
   \caption{Stacked metallicity profile as function of radius of cool (green) and warm/hot (red) gas in the VELA (top) and the NIHAO (bottom) runs. We show results at $z=2.2$ (left), 1.0 (middle) and 0.0 (right). Solid lines show mean values. Shadowed region show 1$\sigma$ dispersion of the mean. Horizontal dashed black lines are eye-guiding lines at solar metallicity.}
   \label{fig:6.5}%
\end{figure*}

To know the metallicity distribution in the CGM is important in order to determine the observability of specific ion lines and to understand the gas cooling process. Metallicity distribution is studied here through the analysis of several figures, and at a large number of time steps that range from $z=4$ down to $z=1$ in the VELA and down to $z=0$ in the NIHAO. Here we show only a sample of them, the ones not shown here are available online or upon request to the authors. Figures shown in this section are: the distributions of gas metallicity and mass in the space of radial velocities $V_{\textrm r}$ versus radius $R/R_{\textrm v}$ (Fig.~\ref{fig:6.3}), the radial metallicity profiles (Fig.~\ref{fig:6.5}), the mass-weighted metallicity in the $T-\rho$ space (left panel of Fig.~\ref{fig:6.1}) and the projected mass-weighted average metallicity in the real space and along arbitrary lines of sight (Fig.~\ref{fig:6.2}). \smallskip \\
Most of the results on metallicity distribution arising from figures presented in this section can be understood under the general picture of galaxy formation and evolution \citep{BirnboimDekel03,DekelBirnboim06,Keres09,Fielding17}. First, redshift variations on star formation rate and SNe feedback strength strongly affect metallicity, both in its distribution on the real space and on its fraction in cold, cool and warm-hot gas phases. Second, expected redshift variations of the inflow rate of cold/cool low-metallicity gas from IGM also have a strong impact on CGM gas metallicity, more important in its outskirts. Processes like the generation of a virial shock and a warm-hot CGM will also affect the circulation of the different gas phases inside the CGM, as so its metallicity. Finally, variations on the implementation of SNe feedback related physical mechanisms drive to the differences observed between results we get from the NIHAO and the VELA suites.

\subsubsection{Metallicity in VELA}
\begin{figure*}
   \centering
  \begin{tabular}{@{}c@{}}
  \vspace{3.35cm}
        \includegraphics[scale=0.1185]{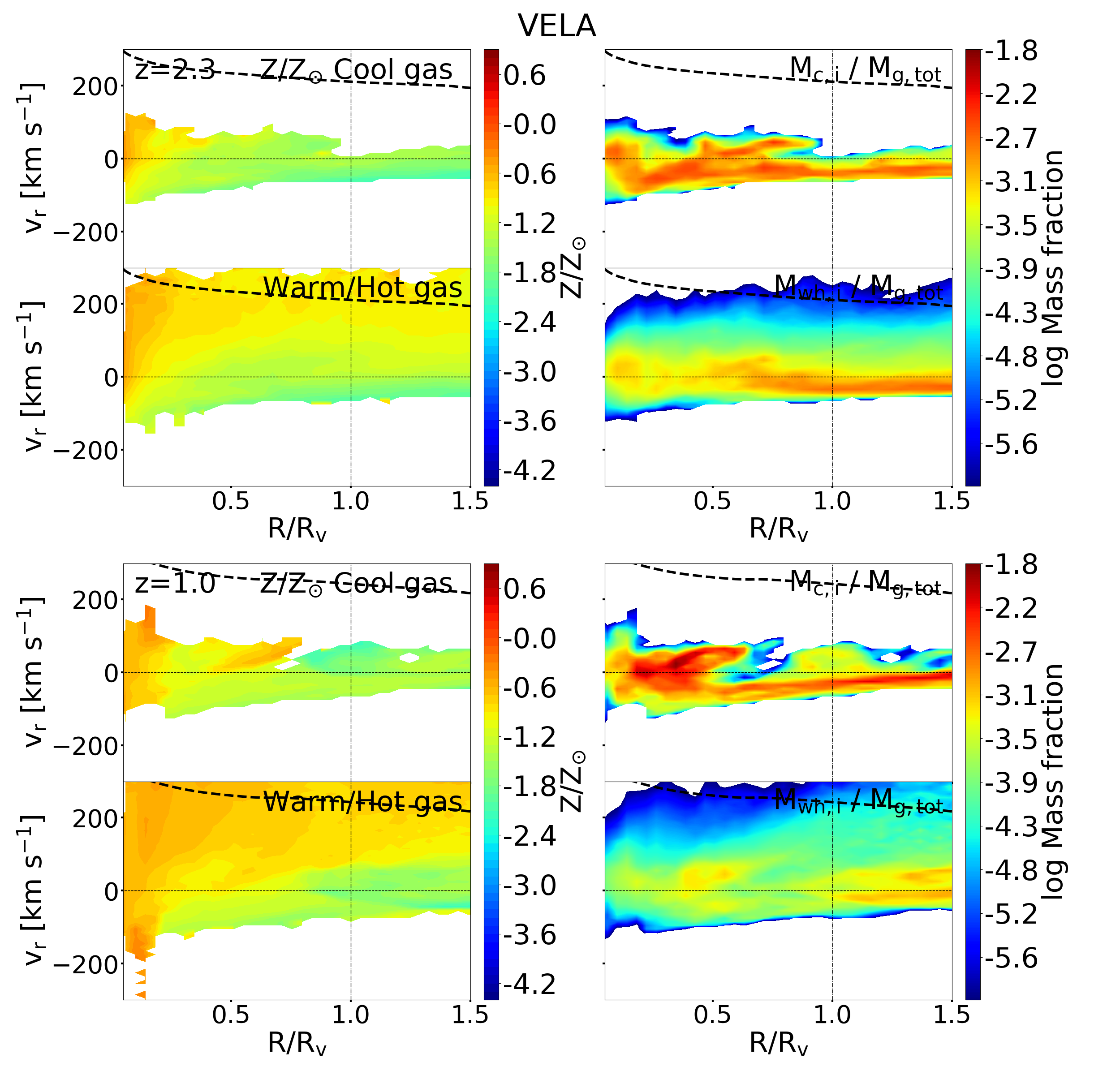}
  \end{tabular}
  \begin{tabular}{@{}c@{}}
        \includegraphics[scale=0.12]{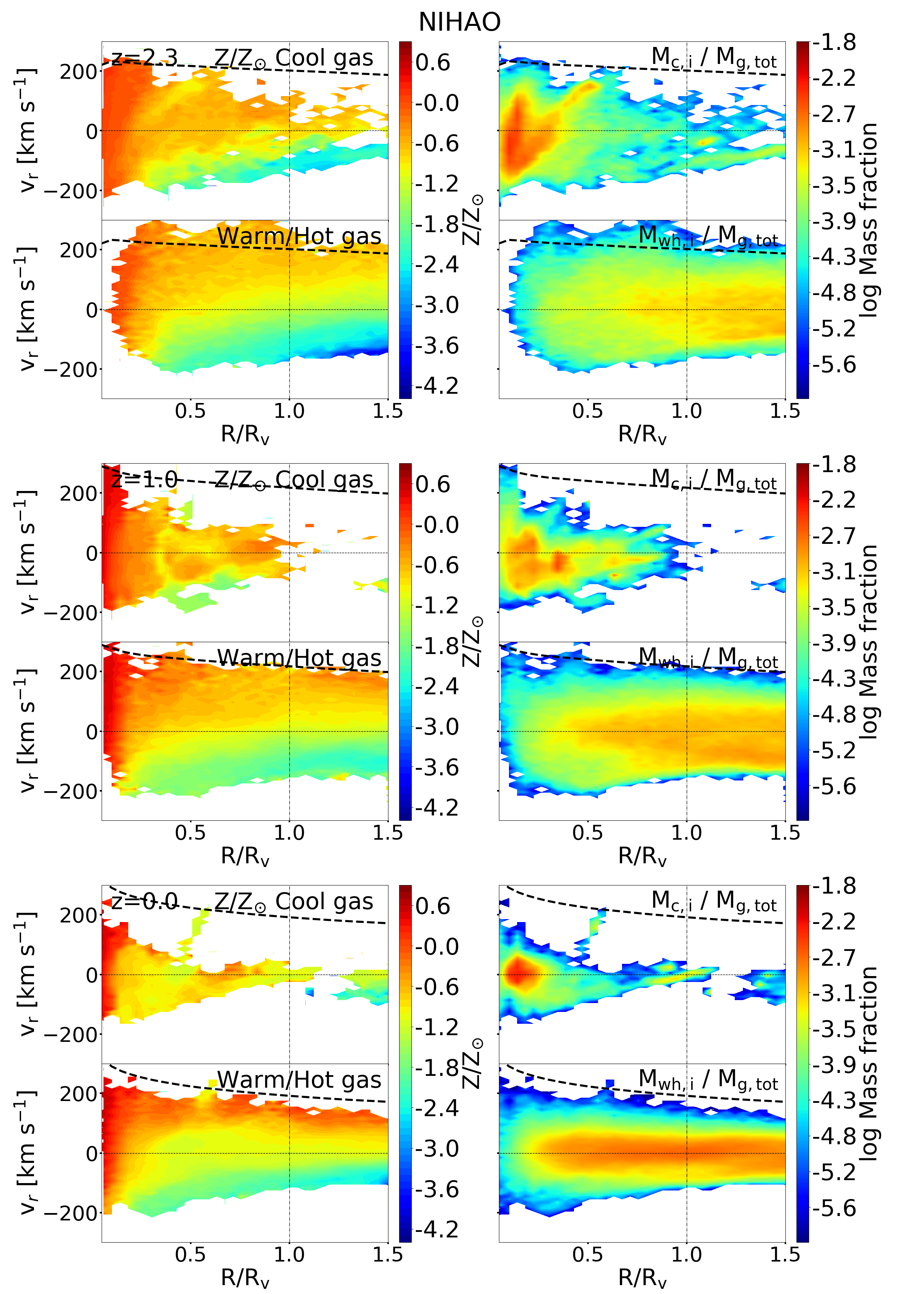}
  \end{tabular}
   \caption{Cool and warm/hot gas $Z/Z_{\odot}$ (first and third columns) and logarithm of the gas mass fraction contained in a 10~km~s$^{-1}\times0.05~R/R_{\textrm v}$ cell (second and last columns), in $V_{\textrm r}-R/R_{\textrm v}$ space. In this figure we show values obtained from stacking the VELA runs (first and second column) at $z=2.3$ (top) and $z=1.0$ (bottom), and NIHAO (third and last columns) at $z=2.3$ (top), $z=1.0$ (middle) and $z=0.0$ (bottom). Horizontal and vertical black dashed lines are guiding lines. Thick black dashed lines show $v_{\textrm esc}$ profile computed using $GM/r$ approximation. }
   \label{fig:6.3}%
\end{figure*}

Figure~\ref{fig:6.5} shows the mean metallicity profile as a function of radius. We find that at all redshifts, at radii larger than $\sim 0.3~R_{\textrm v}$, the metallicity is lower for the cool gas than in the warm/hot gas. Below $\sim 0.3~R_{\textrm v}$ the metallicity is similar for both components, and only below 0.1~$R_{\textrm v}$ is close to solar. The slope of the metallicity profile is marginally redshift dependent; we can see that at high redshift it decreases slightly with radius (outside the disk region), while at $z=1$ this decrease is more evident, with a more negative slope, for the cool gas component. \smallskip \\
In Fig.~\ref{fig:6.3} we see that the cool gas metallicity-enriched structure present at low radius also spans the whole range of radial velocity. This central metallicity-enriched structure evolves with $z$ due to that metallicity enriched cool gas reaches larger radii at lower redshift (see Fig.~\ref{fig:6.3}), following the increased levels of star formation and feedback that peaks at around $z=1.5$ \citep{MadauDickinson14}.  Cool gas with low metallicity and negative radial velocities is seen at large radii, reflecting inflowing cold streams of relatively fresh gas (see Fig.~\ref{fig:6.3}). \smallskip \\
Unlike cool gas, the warm/hot gas phase evolves strongly with $z$. The warm/hot gas phase spans a larger range of radial velocities and metallicities at low-$z$ than at mid-high-$z$ (see Fig.~\ref{fig:6.3}). This is consistent with stronger outflows at low-$z$ that pollute the CGM and generate a warm/hot gas corona. We identify a warm/hot component with low metallicity and low radial velocity (cyan structures in Fig.~\ref{fig:6.3}, second column), which at high-$z$ penetrates to small radii. This might have been pushed inwards by the inflowing cool gas. The kinematics will be discussed further in Sec.~\ref{sec:4.5}. \smallskip \\

\begin{figure}
   \hspace{-0.7cm}
   \includegraphics[scale=0.15]{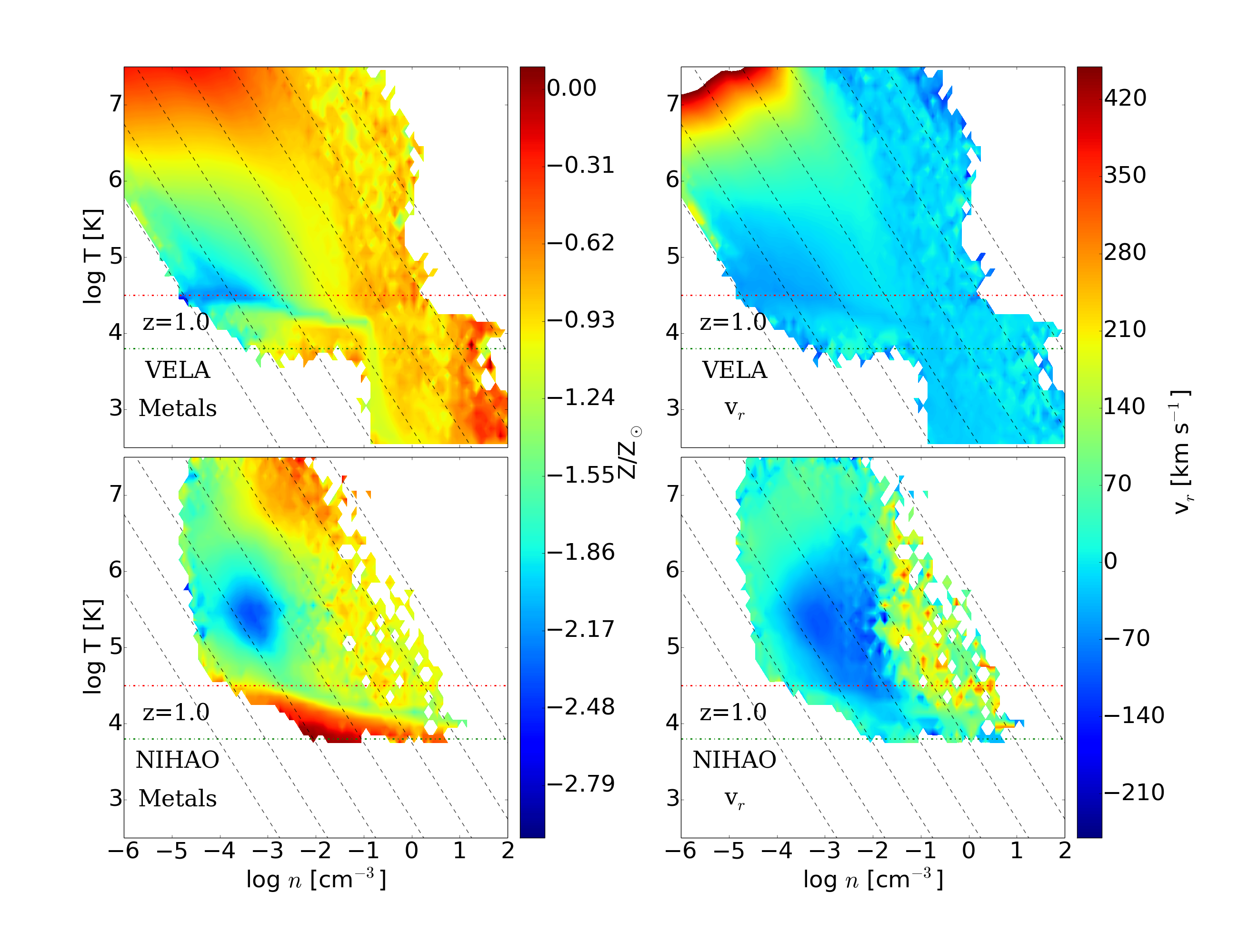}
   \caption{Stacked $Z/Z_{\odot}$ (left) and $V_{\textrm r}$ (right) in the $T-\rho$ space for the VELA simulations (top row) and the NIHAO simulations (bottom row), at $z=1.0$. Positive $V_{\textrm r}$ represents outflowing gas, negative is inflowing. Diagonal gray dashed lines show isobaric evolution. Green and red horizontal dot-dashed lines indicate the transition between cold/cool and cool-warm/hot gas phases, respectively. Only properties of gas in the CGM (i.e., 0.1~$R_{\textrm v}-1.0~R_{\textrm v}$) are shown.}
   \label{fig:6.1}%
\end{figure}

\begin{figure}
   \centering
   \includegraphics[scale=0.17]{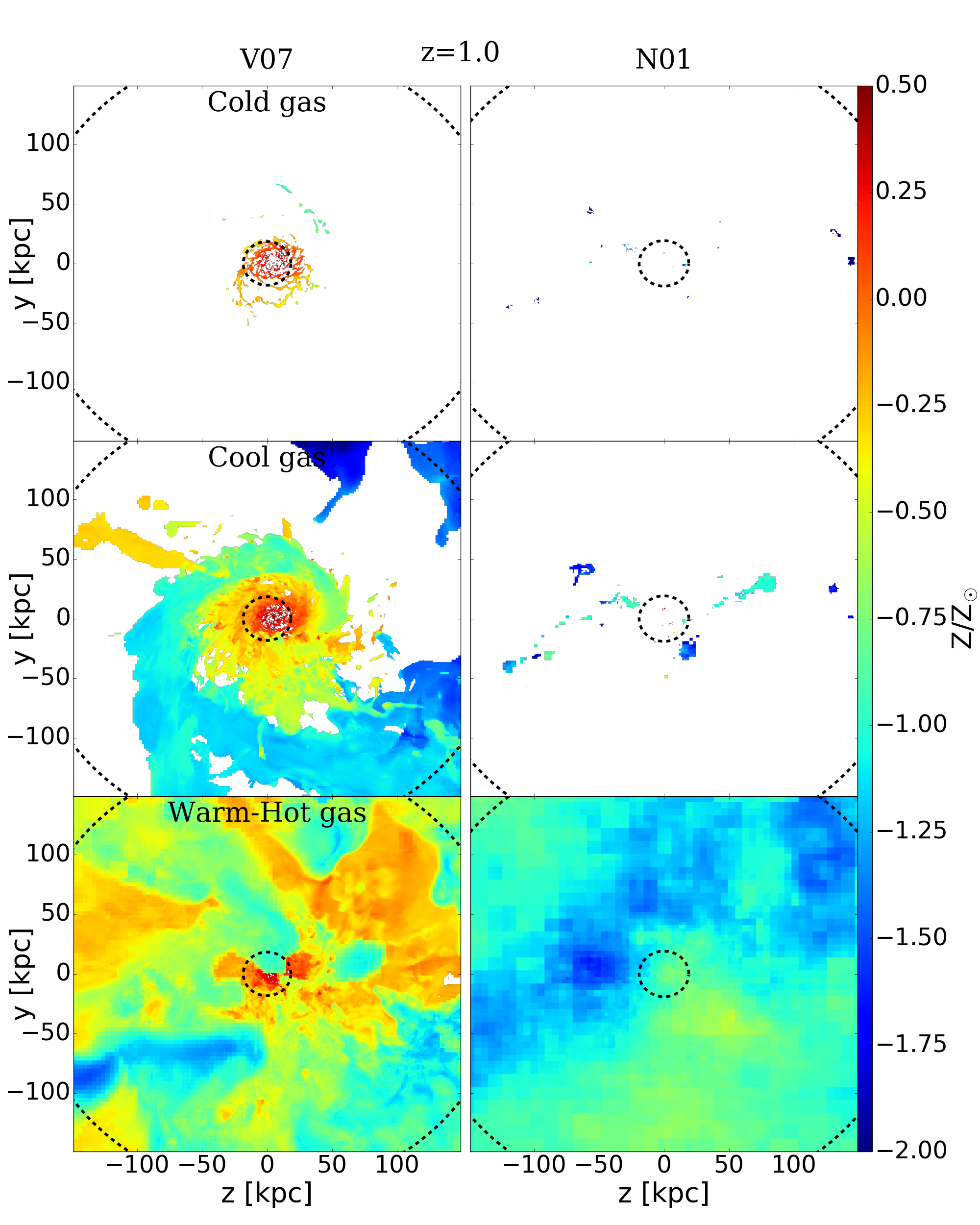}
   \caption{Mass-averaged metallicity along lines of sight perpendicular to the y-z plane. We show metallicity of cold (top), cool (middle) and warm/hot (bottom) gas in V07 (left) and N01 (right). The inner dashed circle indicates the position of 0.1~$R_{\textrm v}$, and the outer circle indicates the virial radius ($R_{\textrm v}$) location. White regions are regions with no gas in the corresponding temperature phase.}
   \label{fig:6.2}%
\end{figure}

In Fig.~\ref{fig:6.1} (top-left panel) we show the VELA stacked $T-\rho$ phase-space diagram colored by metallicity. In this figure one can envision how feedback from the galaxy heats enriched gas from the cold phase into the cool, warm, and finally hot phase, following the lines of constant pressure. One can also envision that low metallicity gas enters the halo from the IGM as gas of low density and low temperature (middle-left region of the $T-\rho$  diagram). Part of this inflowing gas is heated and mixed with the warm/hot gas phase and part inflows into the disk as it cools further. These results have been confirmed by the analysis of the temporal evolution of gas distribution, both in the real space and int the phase-space diagrams\footnote{A larger set of figures is available online. The full set of figures is available upon request to the authors.}. Also the gas origin analysis presented in Sec.~\ref{sec:4.6} supports the conclusions presented here.\smallskip \\
Finally, the maps in Fig.~\ref{fig:6.2} illustrate again, especially for the V07 model (left panels), that the cold component tends to be enriched disk gas. Both the cool and the warm/hot components tend to be metal rich in the central regions, where they are contaminated by supernova feedback. In contrast, both components are more metal poor in the outer regions, where they are inflowing from or mixed with the IGM (see also Fig.~\ref{fig:6.5}). \smallskip \\

\subsubsection{Metallicity in NIHAO}
The global metallicity features in the NIHAO simulations are similar to the ones shown in VELA, with minor differences due to variations in the feedback recipes. In this section we simply describe features in NIHAO figures, a full comparison between NIHAO and VELA metallicity distribution can be found in Sec.~\ref{comparison}.\smallskip \\

In Fig.~\ref{fig:6.5} we see that both cool (green) and warm/hot gas (red) show a similar metallicity profile at high-$z$. Metallicity at high-$z$ decreases with radius, something that is expected in systems with strong inflow from the IGM and low SNe feedback from the disk. At lower-$z$, the mean metallicity is higher and the profile is flatter. Flattening of the metallicity distribution with $z$ is expected if the warm/hot CGM is shielded from cool inflows by virial shocks. In this figure we see that cool gas has higher metallicity than the warm-hot, at low-$z$. This 
can be explained when analysing metallicity and mean radial velocity, together (see Fig.~\ref{fig:6.3}, third and last columns, and \ref{fig:6.1}, bottom panels). First, cool gas is a mixture of infalling low-metallicity gas in clumps and higher-metallicity gas infalling from IGM and/or cooling down from a static/outflowing hotter phase (see Fig.~\ref{fig:6.3}, third and last columns); at low-$z$ inflows are less efficient at transporting fresh low-metallicity gas to the inner regions, and as a consequence the mean cool gas metallicity is higher.\\
Fig.~\ref{fig:6.3}, last column, illustrates that at high redshift the cool gas component is more massive and resides at a lower radius compared to the warm/hot gas. We observe that at low-$z$ the warm/hot gas mass increases and also becomes more uniformly distributed in radius throughout the CGM. Warm/hot gas evolution reflects the generation of a warm/hot CGM due to the effects of a virial shock, in addition to the fact that outflowing gas is no longer able to escape to the IGM.\\
Third column in Fig.~\ref{fig:6.3} shows that at high-$z$, when no warm/hot gas corona has been yet developed, the cool gas easily penetrates all the way down to the disk (see cyan low velocity structure in top panel). At lower-$z$, only cool gas in dense clumps/satellites is able to reach disk regions after crossing the warm/hot CGM (middle and bottom panels). The absolute metallicity and temperature of inflowing gas also changes with $z$. At high-$z$ we observe cool gas with low metallicity almost reaching the disk regions. At lower-$z$ there is almost no low-metallicity cool gas penetrating to the CGM region from the IGM (see Sec.~\ref{sec:4.7.2} for a more detailed discussion about cool gas penetration at $R_{\textrm v}$), but most of it is in a warmer phase. In the disk region a metallicity enhanced cool gas component is always present. This cool metallicity enhanced gas is a mixture of infalling gas from the IGM (low metallicity) and recycled high-metallicity gas from previous SN outflows. Finally, as expected, the highest metallicity gas in the CGM is found in the fast warm/hot outflows of SNe (see warm-hot gas at or close to scape velocity). \smallskip \\

\subsubsection{Metallicity in NIHAO vs. in VELA}\label{comparison}

From Fig~\ref{fig:6.5} to \ref{fig:6.2} we characterize metallicty distribution in both VELA and NIHAO suites. Some differences coming from differences on implemented feedback recipes arise.\\
First, metallicity of the cool gas phase clearly differs from one suite to the other (see Fig.~\ref{fig:6.5}). After a deep analysis of the full set of figures from all models in each suite, we conclude that two effects conduct to this difference. The former is that low metallicity gas from IGM in NIHAO is warmer than in VELA (see low metallicity gas in Fig.\ref{fig:6.1}). The latter is cold/cool filaments are denser and thicker in VELA than in NIHAO.\\
A second difference that arises from figures presented here (e.g Fig.\ref{fig:6.2}) is the warm/hot gas distribution. This difference is also due to the different stellar feedback implementation in each suite. In the VELA, gas is heated up and metallicity enhanced by SNe to a very hot gas phase that moves to the CGM and quickly reduces its density. This hot gas cools down in the CGM at roughly constant pressure while mixing with lower-metallicity gas present there from inflows (see top-left region in Fig.\ref{fig:6.1}, top panels). Otherwise, in the NIHAO, gas is heated up to a very dense-hot gas phase (delayed cooling), that is kept in the SNe region (disk), and later released to the CGM where expands (reduces its density) and cools down in a non-isobaric process. \smallskip \\

\subsection{Kinematics}\label{sec:4.5}

In this section we analyze the gas kinematics in the VELA and the NIHAO simulations. Like in the previous section here we show only figures from a few number of time steps, illustrative of the general gas behavior. Statements made here are based on the analysis of a full set of figures that ranges from $z=0$ ($z=1$ in the VELA) to $z=4$.

\subsubsection{Gas kinematics in VELA}

The $V_{\textrm r}-R$ diagrams, colored by mass (Fig.~\ref{fig:6.3}, second column), show that at high-$z$  most of the cool gas has an inward radial velocity, and are flowing towards the disk region.  Positive $V_{\textrm r}$ represents outflowing gas, negative is inflowing. When analyzing these diagrams individually we see that at $z\sim~1$ there is less inflowing cool gas than at higher-$z$. The cool gas at low-$z$ mostly resides in the central region of the CGM and has zero or low outflowing radial velocity. The low-velocity and/or outflowing cool gas is produced by cooling from the warm CGM and/or heating from the cold disk by SNe. The warm/hot gas shows an asymmetric distribution about $V_{\textrm r}=0$ and towards outflowing velocities, at all radii. The warm/hot gas with $V_{\textrm r}<0$ is low-metallicity gas that reaches the disk region only at high-$z$. This component can be produced by mixing with cooler low-metallicity inflows. This interpretation is consistent with the general picture of cold inflows penetrating only to the outer halo at low-$z$, once a warm/hot corona is generated \citep{BirnboimDekel03,DekelBirnboim06,Keres09,Fielding17}. The warm/hot gas velocity distribution shows a clear evolution with redshift. At lower redshift, the outflowing velocities are higher and gas is able to reach escape velocity bringing warm/hot and metallicity enhanced gas to the IGM. At higher redshift, only a small fraction of gas is able to reach escape velocity due to the less intense SN feedback. When analyzing non-stacked figures (not presented in this paper) we also observe inflowing and outflowing metallicity-rich clumps that are confined to specific radii. These structures are minor mergers of gas rich satellites. \smallskip \\
From $V_{\textrm r}$ in the $T-\rho$ plane (Fig.~\ref{fig:6.1}, top), at $z=1$ we learn that the hotter gas has higher outflow velocities. This is gas that was ejected from the disk by supernova feedback, and will either cool, slow down, and remain in the CGM or eventually escape to the IGM. We also observe that the inflowing gas is mainly cool gas in the VELA simulations and somewhat warmer in the NIHAO runs. This warm/cool inflowing gas is a combination of gas from the IGM through cold flows and gas that cools down from the warm/hot phase. \smallskip \\

\subsection{Gas origin}\label{sec:4.6}

\begin{figure*}
   \centering
   \includegraphics[scale=0.29]{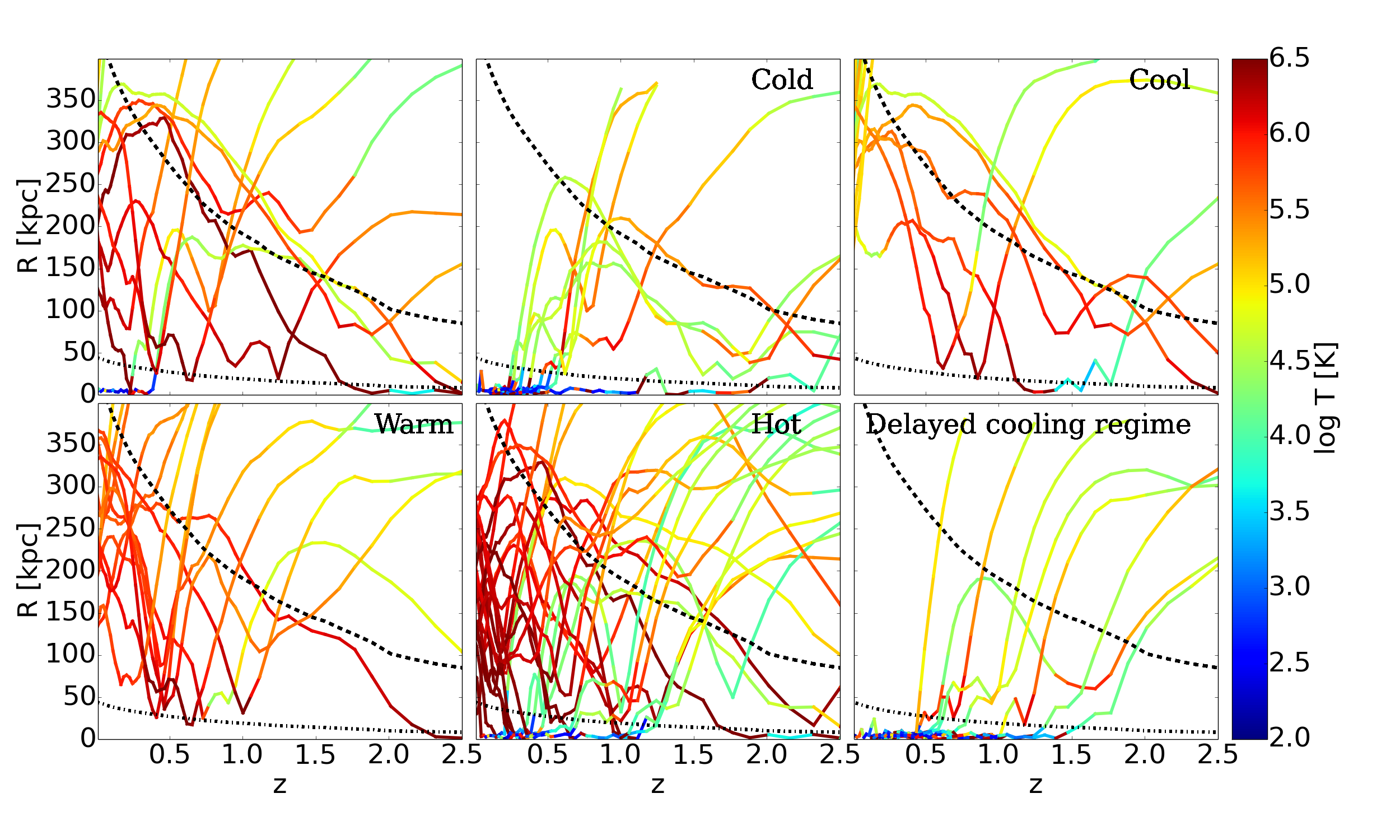}
   \caption{Redshift trajectories of SPH gas particles in the time$-$distance (to the galactic center) space. We show results from the N01 simulation. Each line shows the evolution of a single randomly selected SPH particle. Particles have been randomly selected in multiple Monte-Carlo realizations at $z=0$. Line colors indicate gas particle temperature: blue is cold, green/yellow is cool, orange/light-red is warm and dark-red is hot. We show the evolution of the full sample of particles (top-left panel) and of subsamples that are at cold, cool, warm, hot and delayed cooling phase at z=0 (from top-left to bottom-right panels). The black dashed line shows $R_{\textrm v}$ evolution. The black dot-dashed line indicates the disk region evolution defined as $0.1R_{\textrm v}$. In this figure we can see the different origin of each gas phase at $z=0$. We can also see the temperature evolution and how it depends on its position inside the galactic halo.}
   \label{fig:particles}%
\end{figure*}

In the SPH-type NIHAO simulations it is straightforward to trace the gas particles back in time to reveal their origin and flows in to and out of the CGM. Here we present a first analysis of CGM gas particles evolution that allows us to better understand its metallicity and kinematics evolution. A full analysis of the NIHAO gas particles evolution in time will be presented in a new paper from the NIHAO collaboration. In Fig.~\ref{fig:particles} we show a set of temporal-radial trajectories of randomly selected SPH particles from N01 (equivalent figures from all NIHAO simulations are available upon request to the authors). We show in each panel the radial evolution of particles which end up in each different phase at $z=0$. The thermal evolution of each gas particle is tracked using line colors. Unlike all other plots shown in this paper, here we look at warm and hot gas independently. This decision had been taken in order to find differences on its origin. The results obtained in the analysis of the full set of the NIHAO simulations are as follows: \smallskip \\
\begin{itemize}
\item The cold gas has two different origins: instreaming from the IGM and cooling from the warm CGM. Gas that comes straight from the IGM is heated up by the interaction with the warm/hot CGM; this heating is stronger at low redshift when a warm/hot CGM corona has been produced. Cold gas is mainly located in the galactic disk.
\item The cool gas is either gas from the disk that was heated up by stellar feedback and later on cooled back down or gas that instreams from the IGM/CGM for the first time. This gas spans a large range of radii as it is located both in the disk, infalling through filaments/gas clumps and in the IGM.
\item The warm gas (10$^{4.5}<T<10^6$~K) is almost always gas that was cooled down from hot gas in the CGM or IGM. This gas resides mainly in the CGM. A tiny fraction of this gas has been heated from cool gas to a higher temperature by SN feedback.
\item The hot gas, like the cold and the cool gas,  has a bimodal origin: a fraction comes from cold/cool gas that has been heated up by stellar feedback in the disk and the remaining is CGM/IGM gas that has been heated up at the virial radius by the virial shock. The fraction of gas that has been heated by each of the two mechanisms depends on redshift.
\end{itemize}

For the $\sim10^{12}M_{\odot}$ NIHAO halos, we find that 41\% of the IGM gas that was located within (1$-$5)~$R_{\textrm v}$ at $z=4$ entered the CGM inside $R_{\textrm v}$ by $z=0$. From this gas, 40\% remained in the CGM, 27.5\% returned to the IGM as hot outflows, 12.5\% entered the disk and remained there, and 20\% entered the disk and then returned to the CGM as hot, metal-rich outflows. \smallskip \\
A fraction of 8.1\% of the original IGM gas entered the disk once and was ejected once. Of this gas, 20\% remained in the CGM, 18\% escaped from the halo, mostly at high  $z$ , and 62\% fell back to the disk. From the latter, 44\% remained in the disk till $z=0$. \smallskip \\
A fraction of 2.8\% of the original IGM gas recycled twice (two entries to the disc and two ejections), and less than 1\% recycled three times. \smallskip \\

\section{OVI distribution}\label{sec:4.7}

We now study the spatial, density and temperature distributions of OVI in the NIHAO and VELA simulations. A full comparison of N$_{OVI}$ with COS-Halos and the most recent N$_{OVI}$ observations will be presented in a forthcoming paper (Strawn et al. in prep.). Some of the most relevant results to be presented in the forthcoming paper are: both NIHAO and VELA simulations agree well with COS-Halos N$_{OVI}$ observations at $z\sim~0.5$; most of the OVI in the NIHAO is produced by CI in the warm-hot gas while in the VELA it is by PI in the cool gas. At $z\sim~0$ the NIHAO underpredicts N$_{OVI}$\citep[see ][]{Gutcke17}; there is no VELA data at $z\sim~0$.\smallskip \\
Here we focus on the study of the dominant ionization mechanism as function of radius, redshift and halo mass. To undertake this analysis we need to obtain the OVI fraction produced in simulations. In the following sections we describe the ionization model we use to get the OVI fraction from total oxygen mass. It is important to mention that from VELA simulations only information about alpha (SNe type II) and iron peak elements (SNe type Ia) is provided. As a consequence, we needed to make additional assumptions on the metallicity distribution to get the final OVI mass. So, we assume that all mass from SNe type II is in oxygen. This is not a very strong assumption as oxygen is the most abundant element produced by SNII \citep{Thielemann96}. In NIHAO  the oxygen abundance is automatically obtained in the simulation. Given an oxygen mass we can easily get the OVI fraction using Cloudy models \citep{Ferland98}, assuming a HM12 background radiation field. \smallskip \\

\subsection{OVI metal lines}\label{sec:3}
In this paper we probe the CGM gas via analysis of the OVI ion. The OVI absorption/emission lines may be used to determine the total gas mass inside the virial radius both of external galaxies and the Milky Way. In this context it is important to know that the total gas mass derived from OVI observations is highly dependent of the process that produces oxygen ionization (see Sec.~\ref{sec:1}). Here we summarize the ionization mechanisms that are involved in the production of OVI and describe the ionization model used to obtain OVI abundances from simulations. An extensive discussion about this topic can be found in \citet{SavageSembach91}. \smallskip \\

\subsubsection{Ionization mechanisms}\label{sec:3.1}
As mentioned earlier, the CI and PI mechanisms compete to produce and destroy OVI ions, with the former dominating at high temperatures almost independent of density, and the latter dominating at low temperatures and densities once an adequate ionizing radiation field is present. An ionizing radiation field is included in the current simulations as a uniform metagalactic UV-background \citet{HaardtMadau12}, with the field intensity varying across cosmic times. VELA simulations include partial self-shielding to the UV radiation, in the densest regions. This partial self-shielding does not have a direct impact on the global properties of the CGM analyzed in this paper as the required density conditions mostly occur in the disk region and only deep inside dense filaments/clumps in the CGM. \smallskip \\

\subsubsection{Ionization model (Cloudy)}\label{sec:3.2}
The production of ions is not implemented during processing of current simulations. For our analysis, we used Cloudy \citep{Ferland98} to obtain the ion fractions of OV, OVI and OVII as a function of gas temperature, density and redshift. In addition to default Cloudy heating and cooling parameters, in models used in this work we have set the following ionization parameters: the coronal gas has been modeled as optically thin, an approximation that is valid at the low volume and column densities of the CGM; as a source of ionizing photons we have used a redshift dependent \citet{HaardtMadau12} radiation field (this is also a reasonable assumption as the dominant source of UV ionizing photons far from galactic disks is the metagalactic radiation field); to end with we have added a redshift dependent contribution of CMB and cosmic rays to the UV metagalactic radiation field. To account for the redshift dependence we have generated models with the same parameters but for a grid of redshifts from $z=0$ to $z=5$. Summarizing, we computed ion fractions in a grid of redshift, gas density and temperature for a single thin layer of gas (0.001~pc) embedded into a UV-CMB-cosmic rays metagalactic radiation field (see Fig.~\ref{fig:0}) \smallskip \\

\subsubsection{CI vs. PI dominance in the T-$\rho$ space}\label{sec:3.3}
\begin{figure}
   \centering
      \includegraphics[scale=0.12]{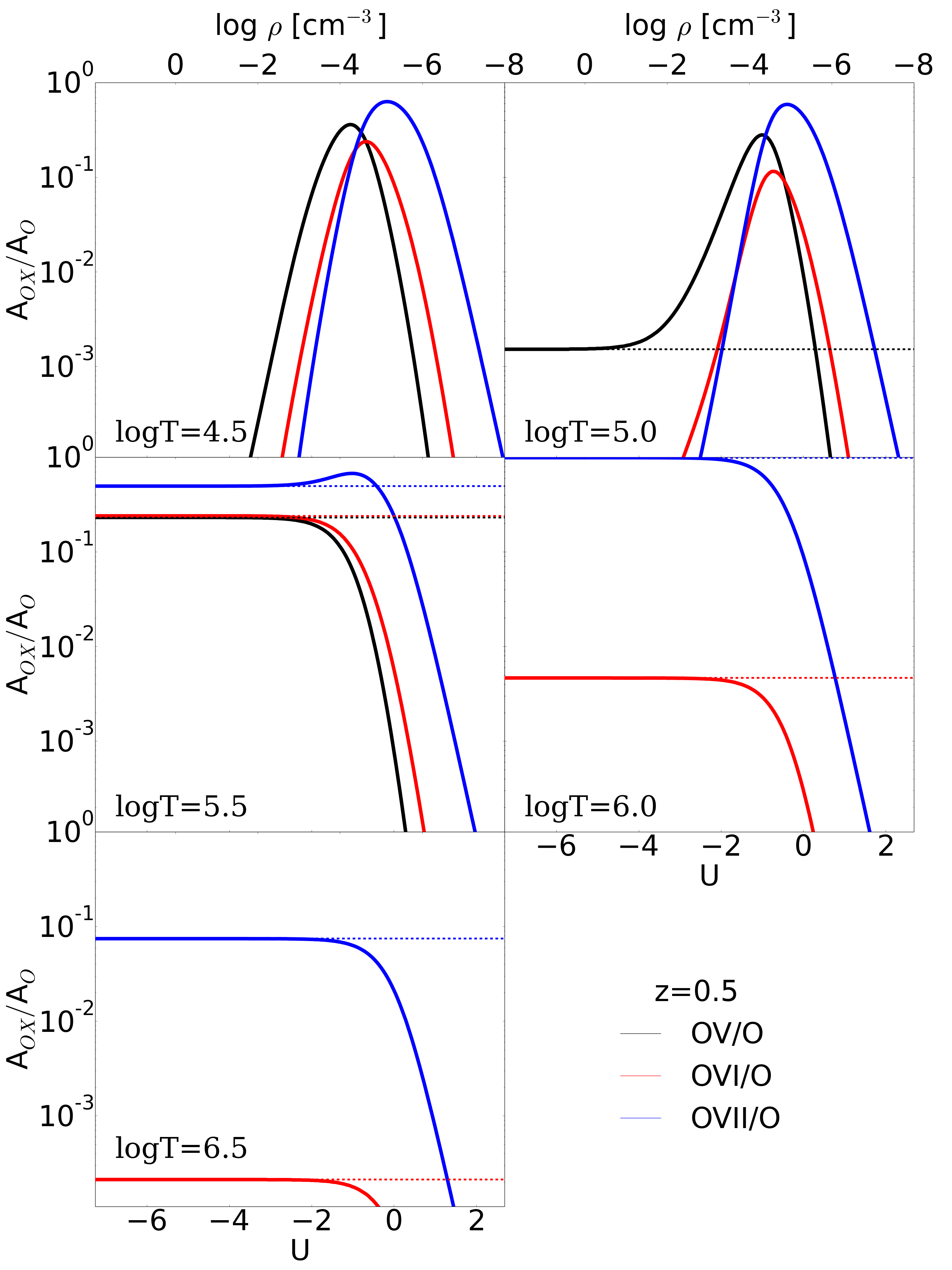}
   \caption{OV (black), OVI (red) and OVII (blue) abundances as function of gas density / ionization parameter U, for metagalactic radiation evaluated at $z=0.5$. Solid lines are results when both collisional ionization and photoionization by a metagalactic UVB are present. Horizontal dashed lines show the oxygen ion abundances if only collisional ionization was accounted. Different panels show the evolution of ionization fraction with gas temperature (see bottom-right labels). Abundances have been obtained by using Cloudy software \citep{Ferland98}.}
   \label{fig:0}%
    \end{figure}

Both, CI and PI, operate together in a wide range of CGM gas temperature and density producing the observed OVI. However, the CI mechanism is more efficient at high temperatures, almost independently of gas density, and PI is almost independent of temperature but highly depends on gas density and the intensity of background radiation field. As a consequence, we can define regions in the $T-\rho$ space where each of the mechanisms is dominant.\\
In Fig.~\ref{fig:0} we show the total ion fractions (both PI and CI-generated) of the OV (black), OVI (red) and OVII (blue) ions as a function of density and temperature at $z=0.5$ (solid lines). In order to better differentiate CI from PI regime we also added the CI-only abundances (dotted lines). From Fig.~\ref{fig:0} it is easily to see that at low temperatures ($T<10^{4.5}$~K) PI is the only relevant ionization mechanism, at a specific density range set by the intensity of the metagalactic radiation field. Regardless of temperature, at lower densities high energy photons are able to produce higher ionization and thereby reduce the abundances of lower ionization states. OVI fraction saturates at a value of about $0.2-0.3$. When temperature increases, collisions start to play a role in ion production, first of OV and later of higher ions. At higher temperatures ($T>10^{5.5}$~K), CI becomes a mechanism to destroy low ionization states in favor of higher ionization. Collisionaly ionized OVI also saturates at a value of $\sim 0.2$. Limits on oxygen ion fractions at low densities are always set by PI. \smallskip \\

\begin{figure*}
   \centering
   \includegraphics[scale=0.28]{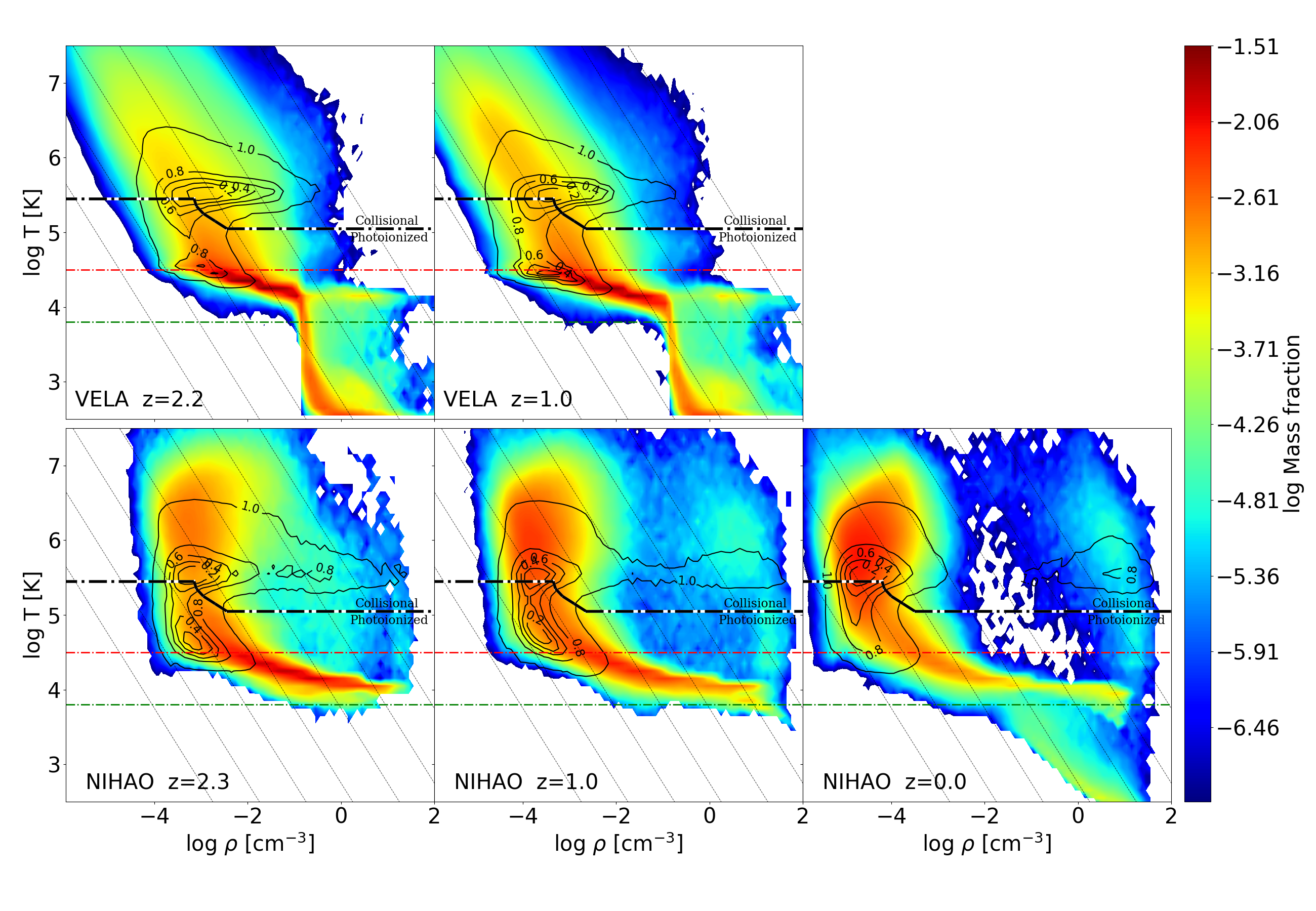}
   \caption{Total gas mass and the OVI mass fractions per bin in $T-\rho$ space. We have used bins of $0.1$~dex both in density and temperature. In color we show the gas mass fraction. Black solid lines are cumulative OVI mass fraction contours. Black-white thick dot-dashed line shows the transition between OVI produced by photoionization (below) and collisional ionization (above). Horizontal dot-dashed lines indicate the transition temperature between different gas phases: cold to cool in green, cool to warm/hot in red. Dashed gray diagonal lines show isobaric evolution. Top panels show stacked results from the VELA set at $z=2.2$ (left) and $1$ (right). Bottom panels show results from the NIHAO set at $z=2.3$ (left), $1$ (middle) and $0$ (right). We only show gas that is in the CGM (i.e., $0.1-1R_{\textrm v}$).}
   \label{fig:10}%
    \end{figure*}

From a set of redshift dependent Cloudy models (e.g. model in Fig.~\ref{fig:0}, at $z=0.5$) we have been able to define the transition from PI dominant (lower temperature) to CI cominant (higher temperature) regimes with the redshift dependent equation shown in Eq.\ref{eq:1}. PI dominates below the curve and CI above. $\rho_z$ is a redshift dependent transition density (see Fig.~\ref{fig:10} for information on values of transition density at $z=0$, $1$ and $2.3$).
\begin{eqfloat*}
\begin{eqnarray}\label{eq:1}
logT(log\rho)=\left\{ \begin{array}{lclcccl}
logT=5.45 & \mbox{if } & &&log\rho&<&log\rho_z  \\
logT=2(log\rho_z-log\rho)+5.45 & \mbox{if }&  log\rho_z&<&log\rho&<&log\rho_z+0.05  \\
logT=0.67(log\rho_z-log\rho)+5.38 & \mbox{if } & log\rho_z+0.05&<&log\rho&<&log\rho_z+0.2  \\
logT=0.4(log\rho_z-log\rho)+5.33 & \mbox{if } &  log\rho_z+0.2&<&log\rho&<&log\rho_z+0.7  \\
logT=5.05 & \mbox{if } & log\rho_z+0.7&<&log\rho&& \end{array}\right.
\end{eqnarray}
\end{eqfloat*}

\begin{figure}
   \centering
      \includegraphics[scale=0.23]{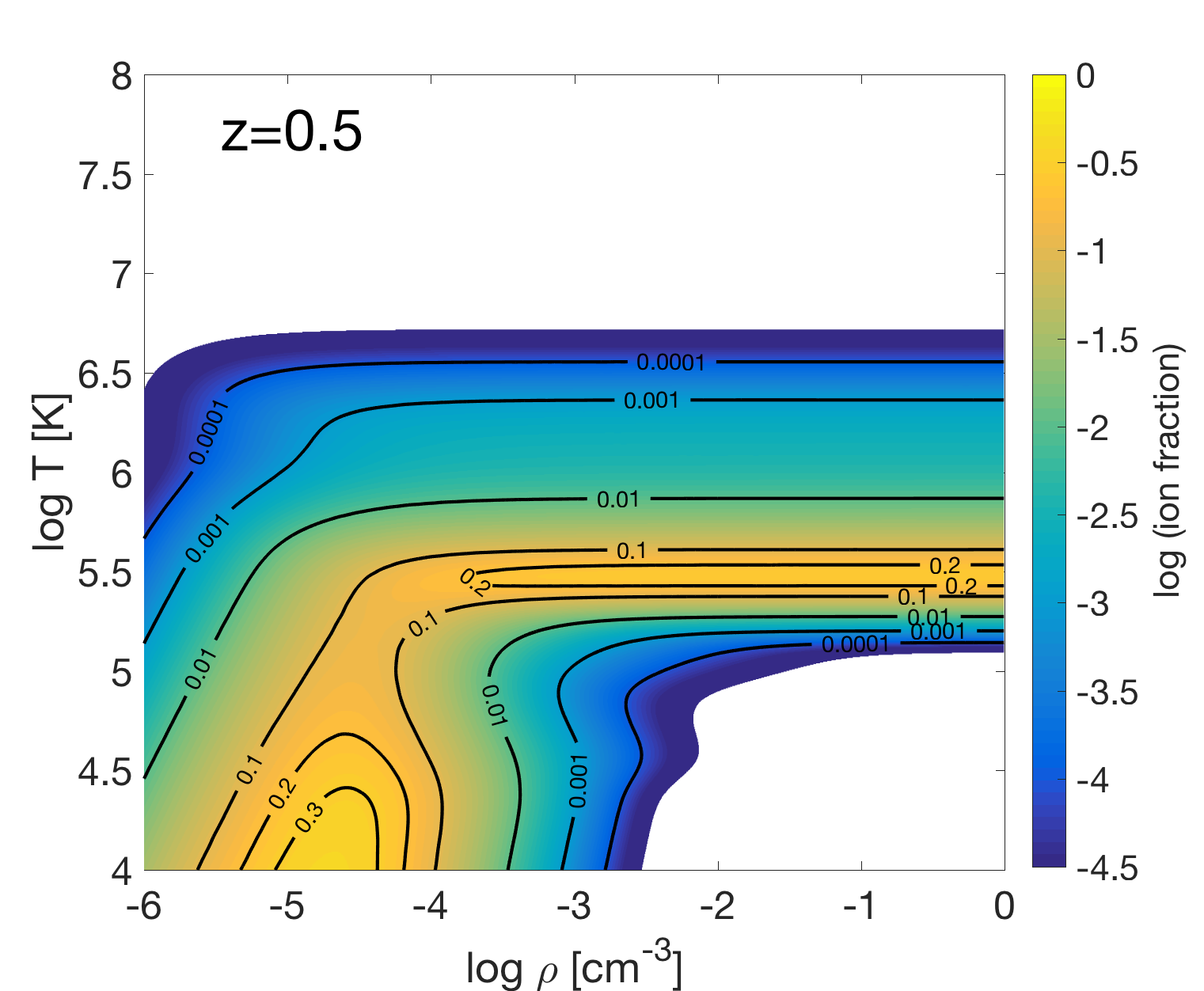}
   \caption{OVI production as function of mass, temperature, and gas density computed using Cloudy 17.00 and assuming an HM12 radiation field, at $z=0$. In color we show the OVI mass fraction. Black solid lines are OVI mass fraction contours.}
   \label{fig:0_1}%
    \end{figure}

In Fig.~\ref{fig:0} we have demonstrated how we define the origin of the OVI ion - CI vs. PI - in the $T-\rho$ parameter space. In Fig.~\ref{fig:0_1} we plot the actual fraction of OVI in this parameter space, which we use to find the OVI mass in our simulation (see results in Fig.\ref{fig:10}.)

\subsubsection{Gas kinematics in NIHAO}

The $V_{\textrm r}-R$ diagrams from the NIHAO simulations (Fig.~\ref{fig:6.3}, third and last columns) clearly show how cool gas enters  (negative $V_{\textrm r}$) to the CGM through filaments or clumps. These cool flows are more clumpy in NIHAO than in VELA. Like in VELA, inflows are more efficient at high-$z$  at bringing the cool gas to the center of the galaxy. The warm/hot outflows reach lower radial velocities at lower redshift; at $z=0$ gas cannot escape and is kept within the CGM. Energy and metals from the low-$z$  outflows remain in the CGM and generate the warm/hot corona. \smallskip \\
From $V_{\textrm r}$ in the $T-\rho$ space (Fig.~\ref{fig:6.1}, bottom), at $z=1$ differences between the feedback recipes in VELA and NIHAO can be seen. The basic difference is that in the NIHAO simulations there is no hot/low-density outflowing gas as it is in the delayed cooling regime and very close to the disk. On the other hand, kinematics of the warm and cool gas are similar in both the VELA and the NIHAO. \smallskip \\

    \begin{figure*}
   \centering
   \includegraphics[scale=0.28]{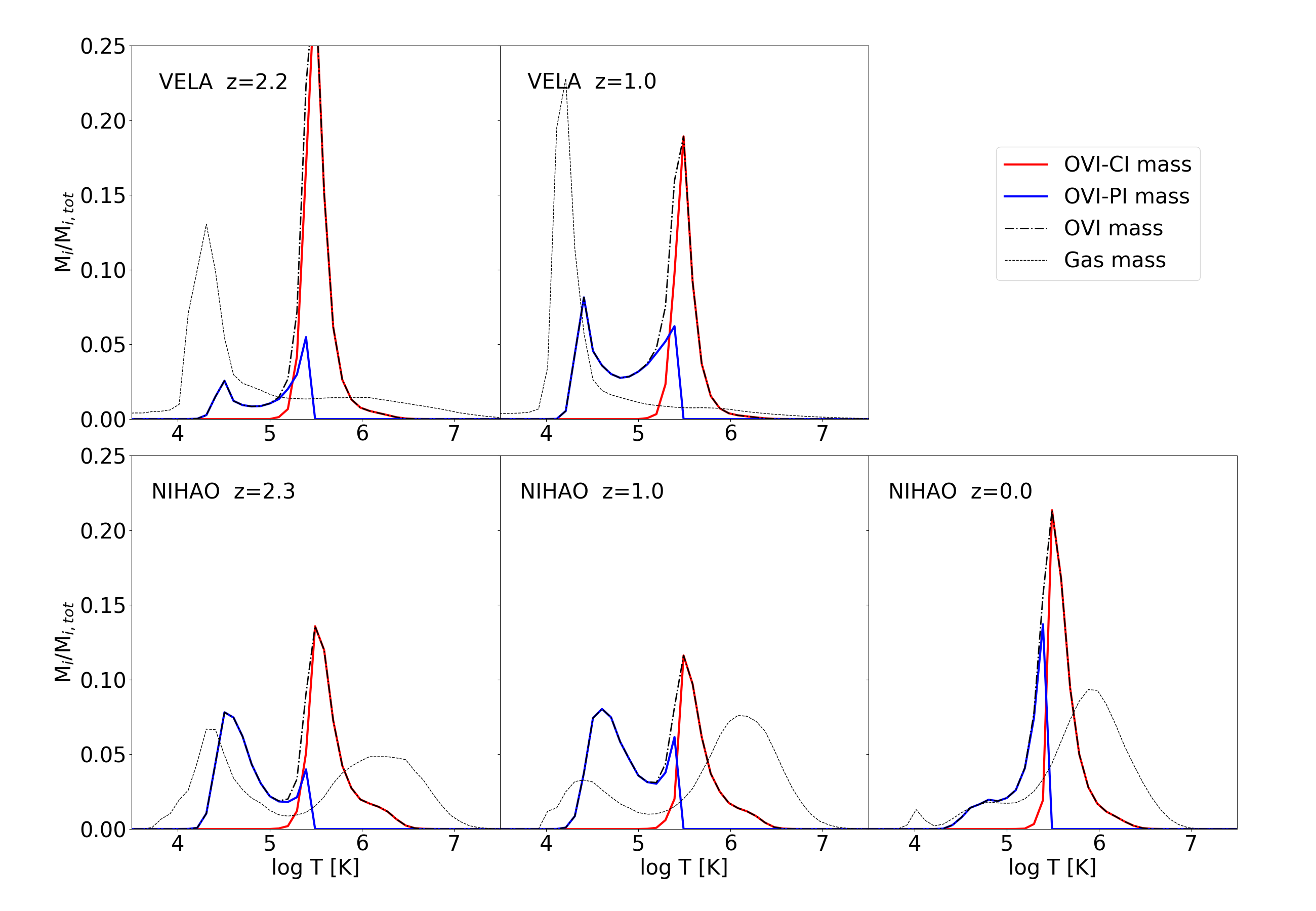}
   \caption{Normalized total gas mass (grey-dotted), total OVI mass (black-dashed) and OVI-CI (red-solid) / OVI-PI (blue-solid) fractions, as function of temperature. Top panels show stacked results from the VELA set at $z=2.2$ (left) and $1$ (right). Bottom panels show results from the NIHAO set at $z=2.3$ (left), $1$ (middle) and $0$ (right). We only show gas that is in the CGM (i.e., $0.1-1~R_{\textrm v}$).}
   \label{fig:Last}%
    \end{figure*} 

\subsection{OVI in the $T-\rho$ space}\label{sec:4.7.0}
Fig.~\ref{fig:10} shows the total gas mass and the OVI gas mass fractions per bin in the $T-\rho$ space. Fig.~\ref{fig:Last} has been generated by collapsing density information in Fig.~\ref{fig:10} over the temperature axes. As a result we obtain the total gas mass, total OVI mass and OVI-PI/OVI-CI mass as function of temperature. We have normalized gas mass per temperature bin by total gas mass (grey-dotted line), and OVI-all (black-dashed) / OVI-PI and OVI-CI (blue and red solid lines, respectively) by total OVI mass.  From Figs.~\ref{fig:10} and \ref{fig:Last} we can confirm that the two mechanisms of oxygen ionization coexist at all redshift in both models. As seen in Fig.~\ref{fig:Last} collisional ionization dominates at $T \gtrsim 3\times 10^5$~K. In Fig.~\ref{fig:10} we see that CI dominance at high temperatures spans a large range of densities. Photoionization dominates in warm or cool gas ($T < 3\times 10^5$~K), in the low density region where the ionization parameter is high enough. The dominant mechanism depends on redshift, both in the NIHAO and the VELA suites. \smallskip \\

We note that the feedback dependent structures discussed in Sec.~\ref{sec:4.2}  (cool-cold gas transition, and hot gas from delayed cooling) have no impact in our analysis of  the OVI in the CGM. The cool-cold gas transition takes place at high density gas, where the photoionization parameter is not high enough for OVI to be produced. The delayed cooling gas component corresponds to less than 1\% of the total gas mass within $R_{\textrm v}$ and is mainly located in the disk region ($0.1R_{\textrm v}$). \smallskip \\

    \begin{figure}
   \centering
      \includegraphics[scale=0.29]{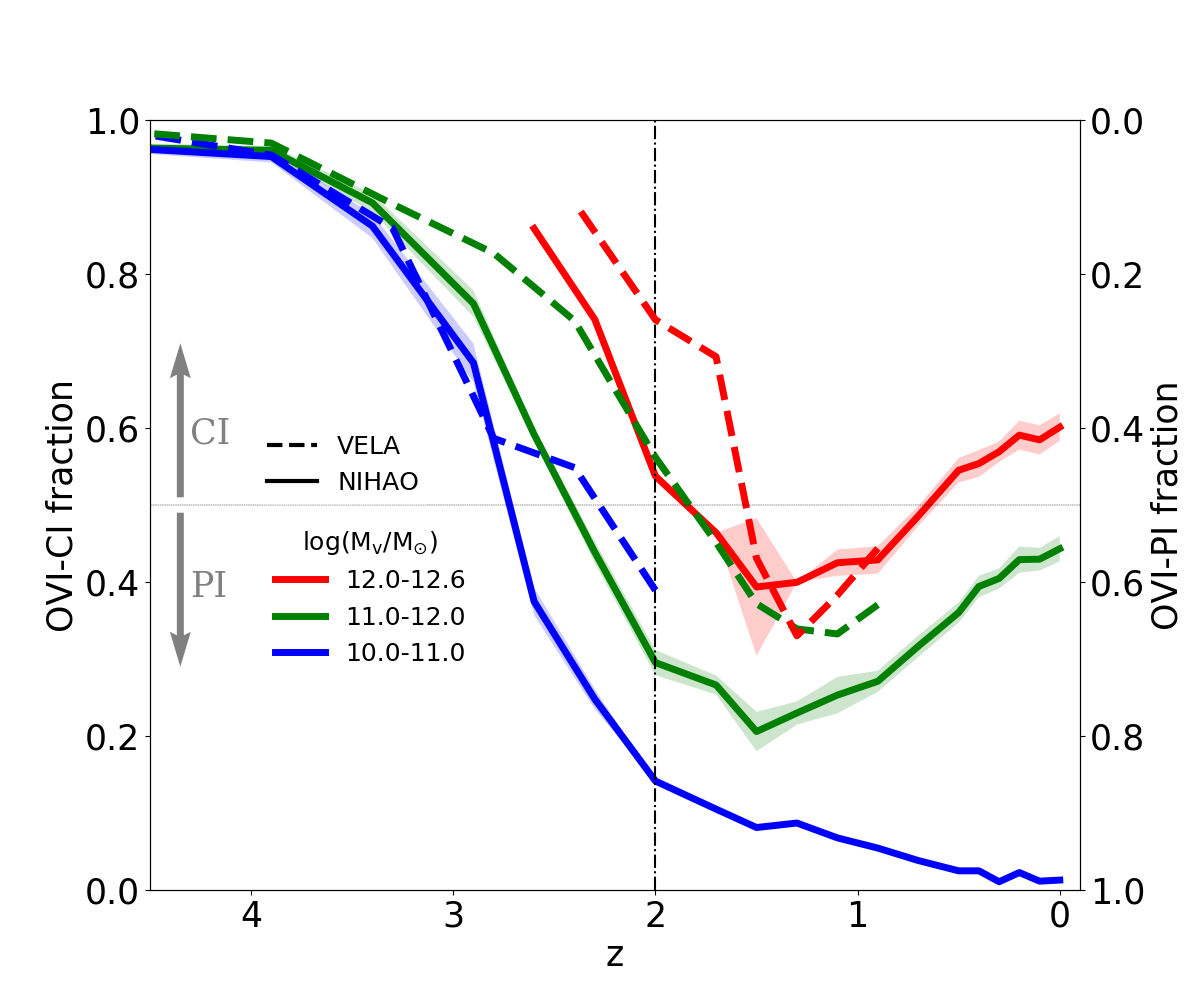}
   \caption{Mass fraction of collisionally ionized OVI as function of redshift. Solid lines show mean values obtained from the NIHAO runs in three different $M_{\textrm v}$ bins (see legend). Dashed lines show the same as solid lines but for the VELA runs.  Dot-dashed vertical black line indicates the peak on the UV-background intensity. Shadowed regions show the 1$\sigma$ dispersion of the mean in the NIHAO results. Only gas in the CGM (i.e. $0.1R_{\textrm v}$ to $R_{\textrm v}$) has been considered.}
   \label{fig:11}
    \end{figure}

\subsection{Redshift-mass dependence}\label{sec:4.7.1}

Fig.~\ref{fig:11} illustrates the collisionally ionized (CI) OVI mass fraction as function of redshift. We show results for fixed halo mass bins at each redshift as opposed to following the galaxies as their mass grows in time. \smallskip \\

\subsubsection{Mass dependence}
The dominant ionization mechanism strongly depends on the halo virial mass. CI is systematically less dominant in lower mass systems at $z<3$. This result is expected as low mass ($M \lesssim 10^{11}~M_{\odot}$) galaxies are surrounded by a cooler CGM that is dominated by inflows through filaments or cold/cool clumps \citep{BirnboimDekel03,DekelBirnboim06,Keres09}. More massive galaxies develop warm/hot coronae through virial shocks. However, at intermediate halo masses, with $M$ in the range of $10^{11}-10^{12}~M_{\odot}$, the CGM may be at temperatures just below the OVI CIE peak, and the metagalactic radiation creates OVI in the warm gas (see $z=0$ panel at Fig. \ref{fig:10}). In the more massive systems ($M>10^{12}~M_{\odot}$), the OVI production is dominated by collisional ionization. At high-$z$ ($z>3$) the mass dependence disappears and CI is the single ionization mechanism. This is a consequence of the low UVB at high-$z$ and the fact that metals are located only in warm/hot gas outflows driven by SNe. At high-$z$ there are almost no metals in the inflowing cool gas as it has not yet been polluted by SNe. This result on the mass dependence of the ionization mechanism agrees with that obtained by \citep{Gutcke17} from NIHAO simulations at $z=0$. \smallskip \\

\subsubsection{Redshift dependence}
For fixed halo mass (colored solid lines in Fig.~\ref{fig:11}) we see that the dominant ionization mechanism depends on redshift. For halos above 10$^{11}M_{\odot}$ the PI fraction peaks at about $z\sim~1.5$. This is a result of both the UV background flux peak at $z\sim~2$, and that at this time cool gas coming from the IGM has already been enriched by metals through galactic outflows. At lower-$z$ the OVI PI production decreases again, following the decrease of the UV-background intensity and due to the inefficiency of cold flows at bringing fresh cool gas into the warm/hot halos of massive galaxies \citep{BirnboimDekel03,Keres09}. For low mass halos ($M<10^{11}~M_{\odot}$) PI remains dominant down to $z=0$. This last situation is a consequence of the fact that small halos never develop a hot CGM and in addition that cool gas at low$-z$ has already been enhanced by metals. \smallskip \\

\subsubsection{Model dependence and missing physical processes}
There is a qualitative agreement in the mass and redshift dependence of the photoionization mechanism in the VELA and the NIHAO suites. However, there is a quantitative difference, which is that the VELA OVI-CI profiles seem to be delayed compared to results from the NIHAO simulations. This difference may be a consequence of using different physical prescriptions (e.g. feedback or self-shielding) in each set of models. We suggest that the redshift of the peak in the OVI PI fraction could be used in the future as a test of the feedback models adopted in simulations. This will only became possible if observations provides us information on CI-PI ratios inside real CGM. \smallskip \\

Some known physical processes are neither included in the simulations or in our analysis. One process that can play a role in the ionization of OVI in the CGM is the emission of galactic UV photons from star forming regions \citep{Cantalupo10,Kannan14,Vasiliev15}. Energetic radiation from the galaxy can contribute to PI in the central regions of the halo \citep[see ][]{Sternberg02}. This will depend on the number and distribution of sources emitting photons with $E>100$~eV, the escape fraction of photons from the galactic disk and the properties of AGN activity \citep[see ][]{Oppenheimer18}. The peak of star formation occurs at $z\sim~1.5$ \citep{MadauDickinson14}, similar to the peak of the metagalactic UVB, and it is higher in high mass galaxies \citep{Wang13}. This new source of photons will enhance the production of OVI through PI specially at the PI peak and in the most massive halos, so it will have a small impact in the general mass-redshift dependence presented in this paper. The addition of this new source of UV photons might also lead to a significant destruction of the OVI in the inner CGM region of galaxies with high star formation rate.
\smallskip \\

\subsection{Radial dependence}\label{sec:4.7.2}

\begin{figure*}
   \centering
      \includegraphics[scale=0.29]{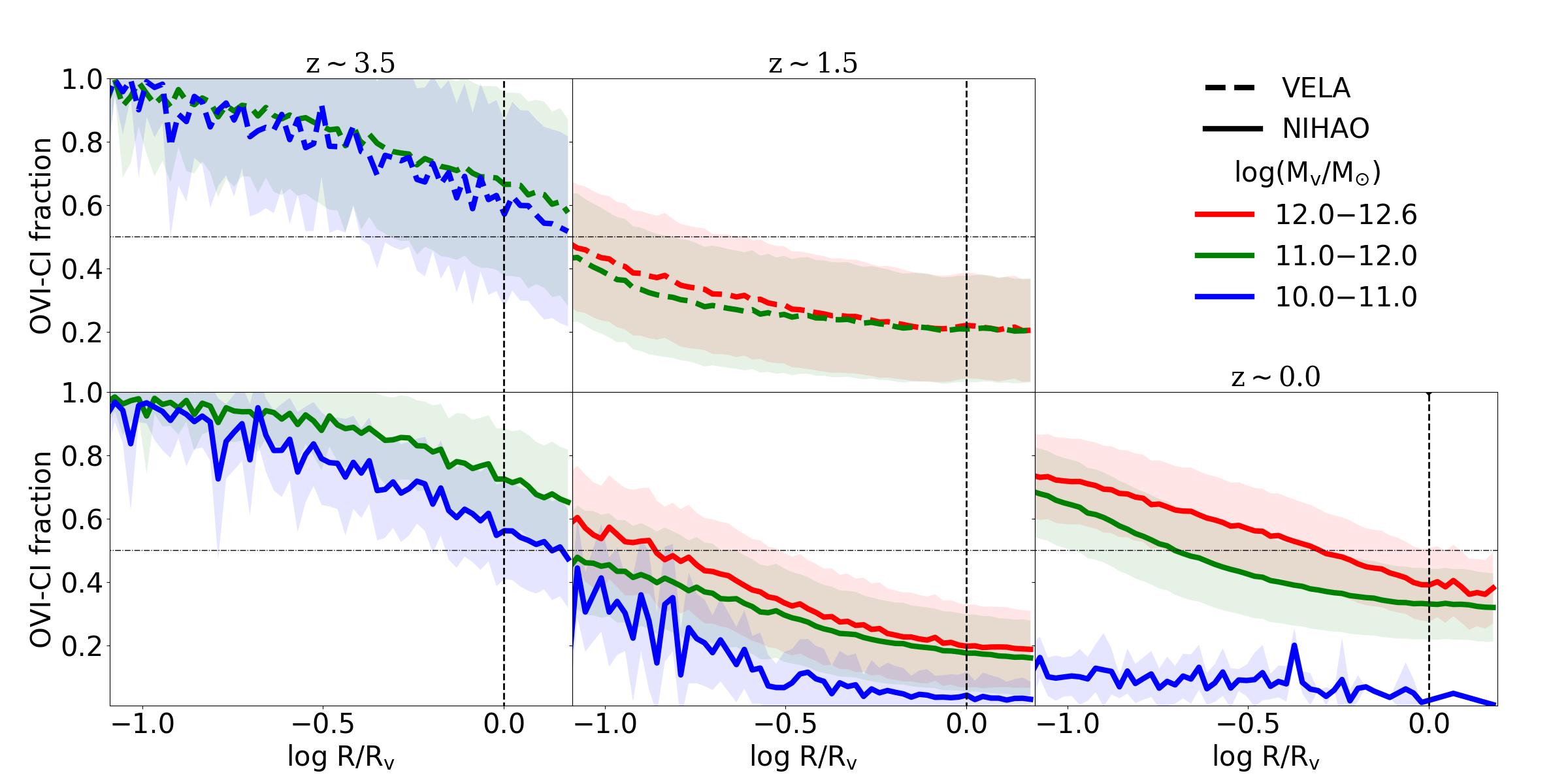}
   \caption{The OVI-CI fraction as function of log $R/R_{\textrm v}$. Top panels: results from the VELA suite at $z\sim~3.5$ (left) and $z\sim~1.5$ (right). Bottom panels: results from the NIHAO suite at  $z\sim~3.5$ (left), $z\sim~1.5$ (middle) and $z\sim~0.0$ (right). Colors show results for three different halo mass bins, i.e, $logM = 10-11$ (blue), $11-12$ (green) and $12-12.6$ (red). Shadowed regions show 1$\sigma$ dispersion from the mean. Dispersion of the mean at each radius in all individual halos and dispersion on the stacking, among different halos, are both accounted. Vertical black-dashed line indicates the position of $R_{\textrm v}$. Horizontal dot-dashed line is a eye-guiding line at the OVI CI/PI co-dominance.}
   \label{fig:11b}
    \end{figure*}

In Fig.~\ref{fig:11b} we show the OVI-CI mass fraction (as defined in Fig.~\ref{fig:11}), as a function of log $R/R_{\textrm v}$. We show the results in three different redshift bins according to the general PI/CI regimes observed in Fig.~\ref{fig:11}. We have found that at larger radii, at all redshift and halo masses studied here, PI has a higher contribution to the OVI production than at smaller radii. It is at large radii where we expect warm/hot gas at lower densities and temperatures (below the CIE OVI peak temperature of $\sim 3 \times 10^5$~K), or inflows of cool gas. At high-$z$ CI dominates in the inner regions independently of halo mass, and in higher mass systems CI is dominant up to larger radii. This result supports the existence of a correlation between outflow strength (mass ejected per unit time) and halo mass. Stronger outflows (i.e more energetic) are able to bring hot gas and metals from the disk farther inside the CGM and also to pollute the IGM. At intermediate $z$, PI dominance extends from the CGM outskirts down to the inner regions, both in the VELA and the NIHAO suites. At low-$z$ in high-mass systems CI dominates to larger radii due to the presence of a virial shock and a lower cool gas accretion rate  (see e.g. Fig.~\ref{fig:1} and \ref{fig:6.3}). From the Fig.~\ref{fig:11b} and the analysis of $V_{\textrm r}-R$ diagrams from individual NIHAO and VELA halos, at a wide range of redshifts, (e.g. Fig.~\ref{fig:6.3} and ~\ref{fig:6.2}, a larger set of figures is available online, the full set is available upon request to the authors) we have found that although a virial shock is produced, warm-hot CGM does not extend up to $R_{\textrm v}$ in all directions, i.e. the virial shock is not located at a constant R. We have also found that the CI/PI co-dominance at low-$z$ is a consequence of the fact that cool gas is still able to penetrate to $R<R_{\textrm v}$, pushing the virial shock to lower radii in specific locations. From our analysis we have concluded that penetration of cool gas depends on the accretion rate (redshift dependence) and also on the warm-hot CGM temperature and density, i.e, outflows (redshift, halo mass and code dependence). 
 \smallskip \\

\section{Summary and Conclusions}\label{sec:8}

In this work we study properties of CGM gas in two sets of simulations obtained using different codes and stellar feedback recipes, the VELA and the NIHAO suites. We focus our analysis on the mass-redshift dependence of the ionization mechanism that produces the OVI observed in the CGM. We note that collisional ionization of OVI occurs in gas at $T \sim 3 \times 10^5$~K (see Fig.\ref{fig:0_1}). Photoionized OVI can originate both in cool or warm gas, as shown by the black contours in Fig.~\ref{fig:10}.
\smallskip \\
The main results are:
\begin{itemize}
\item {\bf Redshift dependence:} The OVI ionization mechanism depends strongly on redshift. At high-$z$ ($z>2-3$) collisions dominate due to a low flux of the ionizing UV-background and low metallicity of cool gas. At lower redshifts photoionization becomes more important as the UV-background flux peaks at $z\sim~2$, and cool/warm gas inflows from the IGM are enhanced by metals from outflows. At lower redshift, collisional ionization becomes important again, especially for high mass systems. The increase in the CI OVI fraction at low-$z$ is a consequence of both a decrease in the metagalactic UV-background intensity and the lower efficiency of inflows which bring fresh cool gas into the warm/hot gas halos of massive galaxies. Systems with $10^{11}~M_{\odot} < M < 10^{12}~M_{\odot}$, develop a warm CGM and at $z \sim 0$, OVI is created both by collisions and photoionization of the warm gas. Low mass systems ($M < 10^{11}~M_{\odot}$) show a different behavior at low-$z$: they do not develop a hot CGM and in addition cool gas at low-$z$ is already enhanced by metals, so PI remains dominant.
\item {\bf Mass dependence:} We confirm that the dominance of an ionization mechanism also depends on the halo mass. This result was previously found by \citet{Gutcke17} when analysing halos at $z=0$, and in this work we extend this analysis to higher redshift. We find that low mass halos, which do not develop a warm/hot corona ($M < 10^{11}~M_{\odot}$), are dominated by photoionization from $z=1-1.5$ down to $z=0$. In systems with higher halo mass, gas is heated by accretion shocks and a warm/hot CGM is formed. The warm/hot gas corona is able to slow down or stop the accretion of fresh cool/warm gas and also keeps hot metallic outflows in the CGM. As a consequence, higher mass halos show higher dominance of collisional ionization in the CGM at low-$z$.
\item {\bf Radial dependence:} We observe a radial dependence on the dominant OVI ionization mechanism. In general, CI is more significant in the central regions while PI is more dominant in the CGM outskirts. The PI is marginally dominant in the outskirts even in high-mass systems at low-$z$. This result may be explained by the existence of warm gas with $T<3 \times 10^5$~K, or the presence of low-density cool gas inflows that are able to push the virial shock to lower radii. Penetration of inflowing gas will depend on the accretion rate and also on the outflow strength.
\item {\bf Feedback dependence:} Although the general behavior of the dominant ionization mechanism as function of redshift and halo mass is similar in VELA and NIHAO, the radial distribution, location in redshift of the peak photoionization fraction and the absolute values of the OVI collisional ionization fraction will strongly depend on feedback recipes used in simulations. The OVI collisional ionization fraction profile, both in redshift and radius, could be used in the future as a new theoretical/observational constraint on the feedback recipes used in simulations.

\item {\bf Origin of the CGM gas:} We have studied the origin of CGM gas in the NIHAO SPH simulations and we have found that each gas phase shows different origins. Interestingly, warm gas can be produced either by cooling of CGM hot gas, by SN heating of disk cold gas, or by inflowing shock heated IGM gas. Cool gas on the other hand can be fresh gas infalling from the CGM in clumps and through filaments, gas from the disk that was heated up by stellar feedback and later on cooled down again, or gas that was heated by the virial shock and cooled down from the IGM/CGM warm/hot gas for the first time. 
 \end{itemize}
\section*{Acknowledgements}

We would like to thank Amiel Sternberg for his valuable comments which helped to improve the manuscript. This work was partly supported by the grants ISF 124/12, I-CORE Program of the PBC/ISF 1829/12, BSF 2014-273, PICS 2015-18, GIF I-1341-303.7/2016, NSF AST-1405962 and by DFG/DIP grant STE 1869/2-1 GE 625/17-1. SFR acknowledges support from a Spanish postdoctoral fellowship "Ayudas para la atracci\'on del talento investigador. Modalidad 2: j\'ovenes investigadores, financiadas por la Comunidad de Madrid" under grant number 2017-T2/TIC-5592. SRF acknowledges financial support from the Spanish Ministry of Economy and Competitiveness (MINECO) under grant number AYA2016-75808-R. The VELA simulations were performed at the National Energy Research Scientific Computing Center (NERSC) at Lawrence Berkeley National Laboratory, and at NASA Advanced Supercomputing (NAS) at NASA Ames Research Center. DC has been funded by the ERC Advanced Grant, STARLIGHT: Formation of the First Stars (project number 339177). NIHAO simulations  were carried out on the High Performance Computing resources at New York University Abu Dhabi; on the theo cluster of the Max-Planck-Institut fuer Astronomie and on the hydra clusters at the Rechenzentrum in Garching.  \smallskip \\




\bibliographystyle{mnras}
\bibliography{biblio} 

\begin{thebibliography}{}
\makeatletter
\relax
\def\mn@urlcharsother{\let\do\@makeother \do\$\do\&\do\#\do\^\do\_\do\%\do\~}
\def\mn@doi{\begingroup\mn@urlcharsother \@ifnextchar [ {\mn@doi@}
  {\mn@doi@[]}}
\def\mn@doi@[#1]#2{\def\@tempa{#1}\ifx\@tempa\@empty \href
  {http://dx.doi.org/#2} {doi:#2}\else \href {http://dx.doi.org/#2} {#1}\fi
  \endgroup}
\def\mn@eprint#1#2{\mn@eprint@#1:#2::\@nil}
\def\mn@eprint@arXiv#1{\href {http://arxiv.org/abs/#1} {{\tt arXiv:#1}}}
\def\mn@eprint@dblp#1{\href {http://dblp.uni-trier.de/rec/bibtex/#1.xml}
  {dblp:#1}}
\def\mn@eprint@#1:#2:#3:#4\@nil{\def\@tempa {#1}\def\@tempb {#2}\def\@tempc
  {#3}\ifx \@tempc \@empty \let \@tempc \@tempb \let \@tempb \@tempa \fi \ifx
  \@tempb \@empty \def\@tempb {arXiv}\fi \@ifundefined
  {mn@eprint@\@tempb}{\@tempb:\@tempc}{\expandafter \expandafter \csname
  mn@eprint@\@tempb\endcsname \expandafter{\@tempc}}}

\bibitem[\protect\citeauthoryear{{Anderson} \& {Bregman}}{{Anderson} \&
  {Bregman}}{2011}]{AndersonBregman11}
{Anderson} M.~E.,  {Bregman} J.~N.,  2011, \mn@doi [\apj]
  {10.1088/0004-637X/737/1/22}, \href
  {http://adsabs.harvard.edu/abs/2011ApJ...737...22A} {737, 22}

\bibitem[\protect\citeauthoryear{{Anderson}, {Gaspari}, {White}, {Wang}  \&
  {Dai}}{{Anderson} et~al.}{2015}]{Anderson15}
{Anderson} M.~E.,  {Gaspari} M.,  {White} S.~D.~M.,  {Wang} W.,   {Dai} X.,
  2015, \mn@doi [\mnras] {10.1093/mnras/stv437}, \href
  {http://adsabs.harvard.edu/abs/2015MNRAS.449.3806A} {449, 3806}

\bibitem[\protect\citeauthoryear{{Birnboim} \& {Dekel}}{{Birnboim} \&
  {Dekel}}{2003}]{BirnboimDekel03}
{Birnboim} Y.,  {Dekel} A.,  2003, \mn@doi [\mnras]
  {10.1046/j.1365-8711.2003.06955.x}, \href
  {http://adsabs.harvard.edu/abs/2003MNRAS.345..349B} {345, 349}

\bibitem[\protect\citeauthoryear{{Bogd{\'a}n}, {Forman}, {Kraft}  \&
  {Jones}}{{Bogd{\'a}n} et~al.}{2013}]{Bogdan13}
{Bogd{\'a}n} {\'A}.,  {Forman} W.~R.,  {Kraft} R.~P.,   {Jones} C.,  2013,
  \mn@doi [\apj] {10.1088/0004-637X/772/2/98}, \href
  {http://adsabs.harvard.edu/abs/2013ApJ...772...98B} {772, 98}

\bibitem[\protect\citeauthoryear{{Bordoloi}, {Wagner}, {Heckman}  \&
  {Norman}}{{Bordoloi} et~al.}{2017}]{Bordoloi17}
{Bordoloi} R.,  {Wagner} A.~Y.,  {Heckman} T.~M.,   {Norman} C.~A.,  2017,
  \mn@doi [\apj] {10.3847/1538-4357/aa8e9c}, \href
  {http://adsabs.harvard.edu/abs/2017ApJ...848..122B} {848, 122}

\bibitem[\protect\citeauthoryear{{Cantalupo}}{{Cantalupo}}{2010}]{Cantalupo10}
{Cantalupo} S.,  2010, \mn@doi [\mnras] {10.1111/j.1745-3933.2010.00806.x},
  \href {http://adsabs.harvard.edu/abs/2010MNRAS.403L..16C} {403, L16}

\bibitem[\protect\citeauthoryear{{Ceverino} \& {Klypin}}{{Ceverino} \&
  {Klypin}}{2009}]{CeverinoKlypin09}
{Ceverino} D.,  {Klypin} A.,  2009, \mn@doi [\apj]
  {10.1088/0004-637X/695/1/292}, \href
  {http://adsabs.harvard.edu/abs/2009ApJ...695..292C} {695, 292}

\bibitem[\protect\citeauthoryear{{Ceverino}, {Dekel}  \& {Bournaud}}{{Ceverino}
  et~al.}{2010}]{Ceverino10}
{Ceverino} D.,  {Dekel} A.,   {Bournaud} F.,  2010, \mn@doi [\mnras]
  {10.1111/j.1365-2966.2010.16433.x}, \href
  {http://adsabs.harvard.edu/abs/2010MNRAS.404.2151C} {404, 2151}

\bibitem[\protect\citeauthoryear{{Ceverino}, {Klypin}, {Klimek},
  {Trujillo-Gomez}, {Churchill}, {Primack}  \& {Dekel}}{{Ceverino}
  et~al.}{2014}]{Ceverino14}
{Ceverino} D.,  {Klypin} A.,  {Klimek} E.~S.,  {Trujillo-Gomez} S.,
  {Churchill} C.~W.,  {Primack} J.,   {Dekel} A.,  2014, \mn@doi [\mnras]
  {10.1093/mnras/stu956}, \href
  {http://adsabs.harvard.edu/abs/2014MNRAS.442.1545C} {442, 1545}

\bibitem[\protect\citeauthoryear{{Ceverino}, {Arribas}, {Colina},
  {Rodr{\'{\i}}guez Del Pino}, {Dekel}  \& {Primack}}{{Ceverino}
  et~al.}{2016}]{Ceverino16b}
{Ceverino} D.,  {Arribas} S.,  {Colina} L.,  {Rodr{\'{\i}}guez Del Pino} B.,
  {Dekel} A.,   {Primack} J.,  2016, \mn@doi [\mnras] {10.1093/mnras/stw1195},
  \href {http://adsabs.harvard.edu/abs/2016MNRAS.460.2731C} {460, 2731}

\bibitem[\protect\citeauthoryear{{Correa}, {Schaye}, {Wyithe}, {Duffy},
  {Theuns}, {Crain}  \& {Bower}}{{Correa} et~al.}{2017}]{Correa17}
{Correa} C.~A.,  {Schaye} J.,  {Wyithe} J.~S.~B.,  {Duffy} A.~R.,  {Theuns} T.,
   {Crain} R.~A.,   {Bower} R.,  2017, preprint, \href
  {http://adsabs.harvard.edu/abs/2017arXiv170901938C} {} (\mn@eprint {arXiv}
  {1709.01938})

\bibitem[\protect\citeauthoryear{{Cowie} \& {Songalia}}{{Cowie} \&
  {Songalia}}{1995}]{CowieSongalia95}
{Cowie} L.,  {Songalia} T.,  1995, in {Butler} B.~J.,  {Muhleman} D.~O.,  eds,
  \baas Vol. 27, Bulletin of the American Astronomical Society. p.~1210

\bibitem[\protect\citeauthoryear{{Crain}, {McCarthy}, {Frenk}, {Theuns}  \&
  {Schaye}}{{Crain} et~al.}{2010}]{Crain10}
{Crain} R.~A.,  {McCarthy} I.~G.,  {Frenk} C.~S.,  {Theuns} T.,   {Schaye} J.,
  2010, \mn@doi [\mnras] {10.1111/j.1365-2966.2010.16985.x}, \href
  {http://adsabs.harvard.edu/abs/2010MNRAS.407.1403C} {407, 1403}

\bibitem[\protect\citeauthoryear{{Dekel} \& {Birnboim}}{{Dekel} \&
  {Birnboim}}{2006}]{DekelBirnboim06}
{Dekel} A.,  {Birnboim} Y.,  2006, \mn@doi [\mnras]
  {10.1111/j.1365-2966.2006.10145.x}, \href
  {http://adsabs.harvard.edu/abs/2006MNRAS.368....2D} {368, 2}

\bibitem[\protect\citeauthoryear{{Dutton} et~al.,}{{Dutton}
  et~al.}{2017}]{Dutton17}
{Dutton} A.~A.,  et~al., 2017, \mn@doi [\mnras] {10.1093/mnras/stx458}, \href
  {http://adsabs.harvard.edu/abs/2017MNRAS.467.4937D} {467, 4937}

\bibitem[\protect\citeauthoryear{{Faerman}, {Sternberg}  \& {McKee}}{{Faerman}
  et~al.}{2017}]{Faerman17}
{Faerman} Y.,  {Sternberg} A.,   {McKee} C.~F.,  2017, \mn@doi [\apj]
  {10.3847/1538-4357/835/1/52}, \href
  {http://adsabs.harvard.edu/abs/2017ApJ...835...52F} {835, 52}

\bibitem[\protect\citeauthoryear{{Ferland}, {Korista}, {Verner}, {Ferguson},
  {Kingdon}  \& {Verner}}{{Ferland} et~al.}{1998}]{Ferland98}
{Ferland} G.~J.,  {Korista} K.~T.,  {Verner} D.~A.,  {Ferguson} J.~W.,
  {Kingdon} J.~B.,   {Verner} E.~M.,  1998, \mn@doi [\pasp] {10.1086/316190},
  \href {http://adsabs.harvard.edu/abs/1998PASP..110..761F} {110, 761}

\bibitem[\protect\citeauthoryear{{Fielding}, {Quataert}, {McCourt}  \&
  {Thompson}}{{Fielding} et~al.}{2017}]{Fielding17}
{Fielding} D.,  {Quataert} E.,  {McCourt} M.,   {Thompson} T.~A.,  2017,
  \mn@doi [\mnras] {10.1093/mnras/stw3326}, \href
  {http://adsabs.harvard.edu/abs/2017MNRAS.466.3810F} {466, 3810}

\bibitem[\protect\citeauthoryear{{Ford} et~al.,}{{Ford} et~al.}{2016}]{Ford16}
{Ford} A.~B.,  et~al., 2016, \mn@doi [\mnras] {10.1093/mnras/stw595}, \href
  {http://adsabs.harvard.edu/abs/2016MNRAS.459.1745F} {459, 1745}

\bibitem[\protect\citeauthoryear{{Goerdt}, {Ceverino}, {Dekel}  \&
  {Teyssier}}{{Goerdt} et~al.}{2015}]{GoerdtCeverino15}
{Goerdt} T.,  {Ceverino} D.,  {Dekel} A.,   {Teyssier} R.,  2015, \mn@doi
  [\mnras] {10.1093/mnras/stv2005}, \href
  {http://adsabs.harvard.edu/abs/2015MNRAS.454..637G} {454, 637}

\bibitem[\protect\citeauthoryear{{Gutcke}, {Stinson}, {Macci{\`o}}, {Wang}  \&
  {Dutton}}{{Gutcke} et~al.}{2017}]{Gutcke17}
{Gutcke} T.~A.,  {Stinson} G.~S.,  {Macci{\`o}} A.~V.,  {Wang} L.,   {Dutton}
  A.~A.,  2017, \mn@doi [\mnras] {10.1093/mnras/stw2539}, \href
  {http://adsabs.harvard.edu/abs/2017MNRAS.464.2796G} {464, 2796}

\bibitem[\protect\citeauthoryear{{Haardt} \& {Madau}}{{Haardt} \&
  {Madau}}{2012}]{HaardtMadau12}
{Haardt} F.,  {Madau} P.,  2012, \mn@doi [\apj] {10.1088/0004-637X/746/2/125},
  \href {http://adsabs.harvard.edu/abs/2012ApJ...746..125H} {746, 125}

\bibitem[\protect\citeauthoryear{{Henley} \& {Shelton}}{{Henley} \&
  {Shelton}}{2010}]{HenleyShelton10}
{Henley} D.~B.,  {Shelton} R.~L.,  2010, \mn@doi [\apjs]
  {10.1088/0067-0049/187/2/388}, \href
  {http://adsabs.harvard.edu/abs/2010ApJS..187..388H} {187, 388}

\bibitem[\protect\citeauthoryear{{Henley}, {Shelton}, {Kwak}, {Joung}  \& {Mac
  Low}}{{Henley} et~al.}{2010}]{Henley10}
{Henley} D.~B.,  {Shelton} R.~L.,  {Kwak} K.,  {Joung} M.~R.,   {Mac Low}
  M.-M.,  2010, \mn@doi [\apj] {10.1088/0004-637X/723/1/935}, \href
  {http://adsabs.harvard.edu/abs/2010ApJ...723..935H} {723, 935}

\bibitem[\protect\citeauthoryear{{Hummels}, {Bryan}, {Smith}  \&
  {Turk}}{{Hummels} et~al.}{2013}]{Hummels13}
{Hummels} C.~B.,  {Bryan} G.~L.,  {Smith} B.~D.,   {Turk} M.~J.,  2013, \mn@doi
  [\mnras] {10.1093/mnras/sts702}, \href
  {http://adsabs.harvard.edu/abs/2013MNRAS.430.1548H} {430, 1548}

\bibitem[\protect\citeauthoryear{{Kannan} et~al.,}{{Kannan}
  et~al.}{2014}]{Kannan14}
{Kannan} R.,  et~al., 2014, \mn@doi [\mnras] {10.1093/mnras/stt2098}, \href
  {http://adsabs.harvard.edu/abs/2014MNRAS.437.2882K} {437, 2882}

\bibitem[\protect\citeauthoryear{{Kere{\v s}}, {Katz}, {Weinberg}  \&
  {Dav{\'e}}}{{Kere{\v s}} et~al.}{2005}]{Keres05}
{Kere{\v s}} D.,  {Katz} N.,  {Weinberg} D.~H.,   {Dav{\'e}} R.,  2005, \mn@doi
  [\mnras] {10.1111/j.1365-2966.2005.09451.x}, \href
  {http://adsabs.harvard.edu/abs/2005MNRAS.363....2K} {363, 2}

\bibitem[\protect\citeauthoryear{{Kere{\v s}}, {Katz}, {Fardal}, {Dav{\'e}}  \&
  {Weinberg}}{{Kere{\v s}} et~al.}{2009}]{Keres09}
{Kere{\v s}} D.,  {Katz} N.,  {Fardal} M.,  {Dav{\'e}} R.,   {Weinberg} D.~H.,
  2009, \mn@doi [\mnras] {10.1111/j.1365-2966.2009.14541.x}, \href
  {http://adsabs.harvard.edu/abs/2009MNRAS.395..160K} {395, 160}

\bibitem[\protect\citeauthoryear{{Kravtsov}}{{Kravtsov}}{2003}]{Kravtsov03}
{Kravtsov} A.~V.,  2003, \mn@doi [\apj] {10.1086/376674}, \href
  {http://adsabs.harvard.edu/abs/2003ApJ...590L...1K} {590, L1}

\bibitem[\protect\citeauthoryear{{Kravtsov}, {Klypin}  \&
  {Khokhlov}}{{Kravtsov} et~al.}{1997}]{Kravtsov97}
{Kravtsov} A.~V.,  {Klypin} A.~A.,   {Khokhlov} A.~M.,  1997, \mn@doi [ApJS]
  {10.1086/313015}, \href {http://adsabs.harvard.edu/abs/1997ApJS..111...73K}
  {111, 73}

\bibitem[\protect\citeauthoryear{{Lehner}, {O'Meara}, {Fox}, {Howk},
  {Prochaska}, {Burns}  \& {Armstrong}}{{Lehner} et~al.}{2014}]{Lehner14}
{Lehner} N.,  {O'Meara} J.~M.,  {Fox} A.~J.,  {Howk} J.~C.,  {Prochaska} J.~X.,
   {Burns} V.,   {Armstrong} A.~A.,  2014, \mn@doi [\apj]
  {10.1088/0004-637X/788/2/119}, \href
  {http://adsabs.harvard.edu/abs/2014ApJ...788..119L} {788, 119}

\bibitem[\protect\citeauthoryear{{Liang}, {Kravtsov}  \& {Agertz}}{{Liang}
  et~al.}{2016}]{Liang16}
{Liang} C.~J.,  {Kravtsov} A.~V.,   {Agertz} O.,  2016, \mn@doi [\mnras]
  {10.1093/mnras/stw375}, \href
  {http://adsabs.harvard.edu/abs/2016MNRAS.458.1164L} {458, 1164}

\bibitem[\protect\citeauthoryear{{Liang}, {Kravtsov}  \& {Agertz}}{{Liang}
  et~al.}{2018}]{Liang18}
{Liang} C.~J.,  {Kravtsov} A.~V.,   {Agertz} O.,  2018, \mn@doi [\mnras]
  {10.1093/mnras/sty1668}, \href
  {http://adsabs.harvard.edu/abs/2018MNRAS.479.1822L} {479, 1822}

\bibitem[\protect\citeauthoryear{{Madau} \& {Dickinson}}{{Madau} \&
  {Dickinson}}{2014}]{MadauDickinson14}
{Madau} P.,  {Dickinson} M.,  2014, \mn@doi [\araa]
  {10.1146/annurev-astro-081811-125615}, \href
  {http://adsabs.harvard.edu/abs/2014ARA%26A..52..415M} {52, 415}

\bibitem[\protect\citeauthoryear{{Mathews} \& {Prochaska}}{{Mathews} \&
  {Prochaska}}{2017}]{Mathews17}
{Mathews} W.~G.,  {Prochaska} J.~X.,  2017, \mn@doi [\apjl]
  {10.3847/2041-8213/aa8861}, \href
  {http://adsabs.harvard.edu/abs/2017ApJ...846L..24M} {846, L24}

\bibitem[\protect\citeauthoryear{{Nelson} et~al.,}{{Nelson}
  et~al.}{2018}]{Nelson18}
{Nelson} D.,  et~al., 2018, \mn@doi [\mnras] {10.1093/mnras/sty656}, \href
  {http://adsabs.harvard.edu/abs/2018MNRAS.477..450N} {477, 450}

\bibitem[\protect\citeauthoryear{{Nicastro} et~al.,}{{Nicastro}
  et~al.}{2002}]{Nicastro02}
{Nicastro} F.,  et~al., 2002, \mn@doi [\apj] {10.1086/340489}, \href
  {http://adsabs.harvard.edu/abs/2002ApJ...573..157N} {573, 157}

\bibitem[\protect\citeauthoryear{{Oppenheimer} et~al.,}{{Oppenheimer}
  et~al.}{2016}]{Oppenheimer16}
{Oppenheimer} B.~D.,  et~al., 2016, \mn@doi [\mnras] {10.1093/mnras/stw1066},
  \href {http://adsabs.harvard.edu/abs/2016MNRAS.460.2157O} {460, 2157}

\bibitem[\protect\citeauthoryear{{Oppenheimer}, {Segers}, {Schaye}, {Richings}
  \& {Crain}}{{Oppenheimer} et~al.}{2018}]{Oppenheimer18}
{Oppenheimer} B.~D.,  {Segers} M.,  {Schaye} J.,  {Richings} A.~J.,   {Crain}
  R.~A.,  2018, \mn@doi [\mnras] {10.1093/mnras/stx2967}, \href
  {http://adsabs.harvard.edu/abs/2018MNRAS.474.4740O} {474, 4740}

\bibitem[\protect\citeauthoryear{{Pachat}, {Narayanan}, {Muzahid}, {Khaire},
  {Srianand}, {Wakker}  \& {Savage}}{{Pachat} et~al.}{2016}]{Pachat16}
{Pachat} S.,  {Narayanan} A.,  {Muzahid} S.,  {Khaire} V.,  {Srianand} R.,
  {Wakker} B.~P.,   {Savage} B.~D.,  2016, \mn@doi [\mnras]
  {10.1093/mnras/stw194}, \href
  {http://adsabs.harvard.edu/abs/2016MNRAS.458..733P} {458, 733}

\bibitem[\protect\citeauthoryear{{Peeples}, {Werk}, {Tumlinson}, {Oppenheimer},
  {Prochaska}, {Katz}  \& {Weinberg}}{{Peeples} et~al.}{2014}]{Peeples14}
{Peeples} M.~S.,  {Werk} J.~K.,  {Tumlinson} J.,  {Oppenheimer} B.~D.,
  {Prochaska} J.~X.,  {Katz} N.,   {Weinberg} D.~H.,  2014, \mn@doi [\apj]
  {10.1088/0004-637X/786/1/54}, \href
  {http://adsabs.harvard.edu/abs/2014ApJ...786...54P} {786, 54}

\bibitem[\protect\citeauthoryear{{Prochaska}, {Weiner}, {Chen}, {Mulchaey}  \&
  {Cooksey}}{{Prochaska} et~al.}{2011}]{Prochaska11}
{Prochaska} J.~X.,  {Weiner} B.,  {Chen} H.-W.,  {Mulchaey} J.,   {Cooksey} K.,
   2011, \mn@doi [\apj] {10.1088/0004-637X/740/2/91}, \href
  {http://adsabs.harvard.edu/abs/2011ApJ...740...91P} {740, 91}

\bibitem[\protect\citeauthoryear{{Rahmati}, {Schaye}, {Crain}, {Oppenheimer},
  {Schaller}  \& {Theuns}}{{Rahmati} et~al.}{2016}]{Rahmati16}
{Rahmati} A.,  {Schaye} J.,  {Crain} R.~A.,  {Oppenheimer} B.~D.,  {Schaller}
  M.,   {Theuns} T.,  2016, \mn@doi [\mnras] {10.1093/mnras/stw453}, \href
  {http://adsabs.harvard.edu/abs/2016MNRAS.459..310R} {459, 310}

\bibitem[\protect\citeauthoryear{{Rasmussen}, {Kahn}  \& {Paerels}}{{Rasmussen}
  et~al.}{2003}]{Rasmussen03}
{Rasmussen} A.,  {Kahn} S.~M.,   {Paerels} F.,  2003, in {Rosenberg} J.~L.,
  {Putman} M.~E.,  eds,  Astrophysics and Space Science Library Vol. 281, The
  IGM/Galaxy Connection. The Distribution of Baryons at z=0. p.~109 (\mn@eprint
  {} {astro-ph/0301183}), \mn@doi{10.1007/978-94-010-0115-1_20}

\bibitem[\protect\citeauthoryear{{Roca-F{\`a}brega}, {Valenzuela},
  {Col{\'{\i}}n}, {Figueras}, {Krongold}, {Vel{\'a}zquez}, {Avila-Reese}  \&
  {Ibarra-Medel}}{{Roca-F{\`a}brega} et~al.}{2016}]{RocaFabrega16}
{Roca-F{\`a}brega} S.,  {Valenzuela} O.,  {Col{\'{\i}}n} P.,  {Figueras} F.,
  {Krongold} Y.,  {Vel{\'a}zquez} H.,  {Avila-Reese} V.,   {Ibarra-Medel} H.,
  2016, \mn@doi [\apj] {10.3847/0004-637X/824/2/94}, \href
  {http://adsabs.harvard.edu/abs/2016ApJ...824...94R} {824, 94}

\bibitem[\protect\citeauthoryear{{Rudie}, {Steidel}, {Shapley}  \&
  {Pettini}}{{Rudie} et~al.}{2013}]{Rudie13}
{Rudie} G.~C.,  {Steidel} C.~C.,  {Shapley} A.~E.,   {Pettini} M.,  2013,
  \mn@doi [\apj] {10.1088/0004-637X/769/2/146}, \href
  {http://adsabs.harvard.edu/abs/2013ApJ...769..146R} {769, 146}

\bibitem[\protect\citeauthoryear{{Savage} \& {Sembach}}{{Savage} \&
  {Sembach}}{1991}]{SavageSembach91}
{Savage} B.~D.,  {Sembach} K.~R.,  1991, \mn@doi [\apj] {10.1086/170498}, \href
  {http://adsabs.harvard.edu/abs/1991ApJ...379..245S} {379, 245}

\bibitem[\protect\citeauthoryear{{Savage}, {Sembach}, {Tripp}  \&
  {Richter}}{{Savage} et~al.}{2002}]{Savage02}
{Savage} B.~D.,  {Sembach} K.~R.,  {Tripp} T.~M.,   {Richter} P.,  2002,
  \mn@doi [\apj] {10.1086/324288}, \href
  {http://adsabs.harvard.edu/abs/2002ApJ...564..631S} {564, 631}

\bibitem[\protect\citeauthoryear{{Shen}, {Wadsley}  \& {Stinson}}{{Shen}
  et~al.}{2010}]{Shen10}
{Shen} S.,  {Wadsley} J.,   {Stinson} G.,  2010, \mn@doi [\mnras]
  {10.1111/j.1365-2966.2010.17047.x}, \href
  {http://adsabs.harvard.edu/abs/2010MNRAS.407.1581S} {407, 1581}

\bibitem[\protect\citeauthoryear{{Steidel}, {Erb}, {Shapley}, {Pettini},
  {Reddy}, {Bogosavljevi{\'c}}, {Rudie}  \& {Rakic}}{{Steidel}
  et~al.}{2010}]{Steidel10}
{Steidel} C.~C.,  {Erb} D.~K.,  {Shapley} A.~E.,  {Pettini} M.,  {Reddy} N.,
  {Bogosavljevi{\'c}} M.,  {Rudie} G.~C.,   {Rakic} O.,  2010, \mn@doi [\apj]
  {10.1088/0004-637X/717/1/289}, \href
  {http://adsabs.harvard.edu/abs/2010ApJ...717..289S} {717, 289}

\bibitem[\protect\citeauthoryear{{Stern}, {Hennawi}, {Prochaska}  \&
  {Werk}}{{Stern} et~al.}{2016}]{Stern16}
{Stern} J.,  {Hennawi} J.~F.,  {Prochaska} J.~X.,   {Werk} J.~K.,  2016,
  \mn@doi [\apj] {10.3847/0004-637X/830/2/87}, \href
  {http://adsabs.harvard.edu/abs/2016ApJ...830...87S} {830, 87}

\bibitem[\protect\citeauthoryear{{Stern}, {Faucher-Gigu{\`e}re}, {Hennawi},
  {Hafen}, {Johnson}  \& {Fielding}}{{Stern} et~al.}{2018}]{Stern18}
{Stern} J.,  {Faucher-Gigu{\`e}re} C.-A.,  {Hennawi} J.~F.,  {Hafen} Z.,
  {Johnson} S.~D.,   {Fielding} D.,  2018, preprint, \href
  {http://adsabs.harvard.edu/abs/2018arXiv180305446S} {} (\mn@eprint {arXiv}
  {1803.05446})

\bibitem[\protect\citeauthoryear{{Sternberg}, {McKee}  \&
  {Wolfire}}{{Sternberg} et~al.}{2002}]{Sternberg02}
{Sternberg} A.,  {McKee} C.~F.,   {Wolfire} M.~G.,  2002, \mn@doi [\apjs]
  {10.1086/343032}, \href {http://adsabs.harvard.edu/abs/2002ApJS..143..419S}
  {143, 419}

\bibitem[\protect\citeauthoryear{{Stinson}, {Seth}, {Katz}, {Wadsley},
  {Governato}  \& {Quinn}}{{Stinson} et~al.}{2006}]{Stinson06}
{Stinson} G.,  {Seth} A.,  {Katz} N.,  {Wadsley} J.,  {Governato} F.,   {Quinn}
  T.,  2006, \mn@doi [\mnras] {10.1111/j.1365-2966.2006.11097.x}, \href
  {http://adsabs.harvard.edu/abs/2006MNRAS.373.1074S} {373, 1074}

\bibitem[\protect\citeauthoryear{{Stinson}, {Brook}, {Macci{\`o}}, {Wadsley},
  {Quinn}  \& {Couchman}}{{Stinson} et~al.}{2013}]{Stinson13}
{Stinson} G.~S.,  {Brook} C.,  {Macci{\`o}} A.~V.,  {Wadsley} J.,  {Quinn}
  T.~R.,   {Couchman} H.~M.~P.,  2013, \mn@doi [\mnras] {10.1093/mnras/sts028},
  \href {http://adsabs.harvard.edu/abs/2013MNRAS.428..129S} {428, 129}

\bibitem[\protect\citeauthoryear{{Stinson} et~al.,}{{Stinson}
  et~al.}{2015}]{Stinson15}
{Stinson} G.~S.,  et~al., 2015, \mn@doi [\mnras] {10.1093/mnras/stv1985}, \href
  {http://adsabs.harvard.edu/abs/2015MNRAS.454.1105S} {454, 1105}

\bibitem[\protect\citeauthoryear{{Stocke}, {Keeney}, {Danforth}, {Shull},
  {Froning}, {Green}, {Penton}  \& {Savage}}{{Stocke} et~al.}{2013}]{Stocke13}
{Stocke} J.~T.,  {Keeney} B.~A.,  {Danforth} C.~W.,  {Shull} J.~M.,  {Froning}
  C.~S.,  {Green} J.~C.,  {Penton} S.~V.,   {Savage} B.~D.,  2013, \mn@doi
  [\apj] {10.1088/0004-637X/763/2/148}, \href
  {http://adsabs.harvard.edu/abs/2013ApJ...763..148S} {763, 148}

\bibitem[\protect\citeauthoryear{{Stocke} et~al.,}{{Stocke}
  et~al.}{2014}]{Stocke14}
{Stocke} J.~T.,  et~al., 2014, \mn@doi [\apj] {10.1088/0004-637X/791/2/128},
  \href {http://adsabs.harvard.edu/abs/2014ApJ...791..128S} {791, 128}

\bibitem[\protect\citeauthoryear{{Suresh}, {Bird}, {Vogelsberger}, {Genel},
  {Torrey}, {Sijacki}, {Springel}  \& {Hernquist}}{{Suresh}
  et~al.}{2015}]{Suresh15}
{Suresh} J.,  {Bird} S.,  {Vogelsberger} M.,  {Genel} S.,  {Torrey} P.,
  {Sijacki} D.,  {Springel} V.,   {Hernquist} L.,  2015, \mn@doi [\mnras]
  {10.1093/mnras/stu2762}, \href
  {http://adsabs.harvard.edu/abs/2015MNRAS.448..895S} {448, 895}

\bibitem[\protect\citeauthoryear{{Suresh}, {Rubin}, {Kannan}, {Werk},
  {Hernquist}  \& {Vogelsberger}}{{Suresh} et~al.}{2017}]{Suresh17}
{Suresh} J.,  {Rubin} K.~H.~R.,  {Kannan} R.,  {Werk} J.~K.,  {Hernquist} L.,
  {Vogelsberger} M.,  2017, \mn@doi [\mnras] {10.1093/mnras/stw2499}, \href
  {http://adsabs.harvard.edu/abs/2017MNRAS.465.2966S} {465, 2966}

\bibitem[\protect\citeauthoryear{{Thielemann}, {Nomoto}  \&
  {Hashimoto}}{{Thielemann} et~al.}{1996}]{Thielemann96}
{Thielemann} F.-K.,  {Nomoto} K.,   {Hashimoto} M.-A.,  1996, \mn@doi [\apj]
  {10.1086/176980}, \href {http://adsabs.harvard.edu/abs/1996ApJ...460..408T}
  {460, 408}

\bibitem[\protect\citeauthoryear{{Thom} \& {Chen}}{{Thom} \&
  {Chen}}{2008}]{ThomChen08}
{Thom} C.,  {Chen} H.-W.,  2008, \mn@doi [\apj] {10.1086/587976}, \href
  {http://adsabs.harvard.edu/abs/2008ApJ...683...22T} {683, 22}

\bibitem[\protect\citeauthoryear{{Tripp}, {Sembach}, {Bowen}, {Savage},
  {Jenkins}, {Lehner}  \& {Richter}}{{Tripp} et~al.}{2008}]{Tripp08}
{Tripp} T.~M.,  {Sembach} K.~R.,  {Bowen} D.~V.,  {Savage} B.~D.,  {Jenkins}
  E.~B.,  {Lehner} N.,   {Richter} P.,  2008, \mn@doi [\apjs] {10.1086/587486},
  \href {http://adsabs.harvard.edu/abs/2008ApJS..177...39T} {177, 39}

\bibitem[\protect\citeauthoryear{{Tumlinson} et~al.,}{{Tumlinson}
  et~al.}{2013}]{Tumlinson13}
{Tumlinson} J.,  et~al., 2013, \mn@doi [\apj] {10.1088/0004-637X/777/1/59},
  \href {http://adsabs.harvard.edu/abs/2013ApJ...777...59T} {777, 59}

\bibitem[\protect\citeauthoryear{{Vasiliev}, {Ryabova}  \&
  {Shchekinov}}{{Vasiliev} et~al.}{2015}]{Vasiliev15}
{Vasiliev} E.~O.,  {Ryabova} M.~V.,   {Shchekinov} Y.~A.,  2015, \mn@doi
  [\mnras] {10.1093/mnras/stu2290}, \href
  {http://adsabs.harvard.edu/abs/2015MNRAS.446.3078V} {446, 3078}

\bibitem[\protect\citeauthoryear{{Wadsley}, {Veeravalli}  \&
  {Couchman}}{{Wadsley} et~al.}{2008}]{Wadsley08}
{Wadsley} J.~W.,  {Veeravalli} G.,   {Couchman} H.~M.~P.,  2008, \mn@doi
  [\mnras] {10.1111/j.1365-2966.2008.13260.x}, \href
  {http://adsabs.harvard.edu/abs/2008MNRAS.387..427W} {387, 427}

\bibitem[\protect\citeauthoryear{{Wadsley}, {Keller}  \& {Quinn}}{{Wadsley}
  et~al.}{2017}]{Wadsley17}
{Wadsley} J.~W.,  {Keller} B.~W.,   {Quinn} T.~R.,  2017, \mn@doi [\mnras]
  {10.1093/mnras/stx1643}, \href
  {http://adsabs.harvard.edu/abs/2017MNRAS.471.2357W} {471, 2357}

\bibitem[\protect\citeauthoryear{{Wang} et~al.,}{{Wang} et~al.}{2013}]{Wang13}
{Wang} L.,  et~al., 2013, \mn@doi [\mnras] {10.1093/mnras/stt190}, \href
  {http://adsabs.harvard.edu/abs/2013MNRAS.431..648W} {431, 648}

\bibitem[\protect\citeauthoryear{{Wang}, {Dutton}, {Stinson}, {Macci{\`o}},
  {Penzo}, {Kang}, {Keller}  \& {Wadsley}}{{Wang} et~al.}{2015}]{Wang15}
{Wang} L.,  {Dutton} A.~A.,  {Stinson} G.~S.,  {Macci{\`o}} A.~V.,  {Penzo} C.,
   {Kang} X.,  {Keller} B.~W.,   {Wadsley} J.,  2015, \mn@doi [\mnras]
  {10.1093/mnras/stv1937}, \href
  {http://adsabs.harvard.edu/abs/2015MNRAS.454...83W} {454, 83}

\bibitem[\protect\citeauthoryear{{Werk} et~al.,}{{Werk} et~al.}{2014}]{Werk14}
{Werk} J.~K.,  et~al., 2014, \mn@doi [\apj] {10.1088/0004-637X/792/1/8}, \href
  {http://adsabs.harvard.edu/abs/2014ApJ...792....8W} {792, 8}

\bibitem[\protect\citeauthoryear{{Zolotov} et~al.,}{{Zolotov}
  et~al.}{2015}]{Zolotov15}
{Zolotov} A.,  et~al., 2015, \mn@doi [\mnras] {10.1093/mnras/stv740}, \href
  {http://adsabs.harvard.edu/abs/2015MNRAS.450.2327Z} {450, 2327}

\makeatother
\end{thebibliography}



\bsp	
\label{lastpage}
\end{document}